\documentstyle[11pt,epsfig]{article}   
\newcommand{\PO}{\em I\! \! P }   
\newcommand{\pom}{I\!\!P}   
\newcommand{\xpom}{x_{\pom}}   
\newcommand{\be}{\begin{equation}}   
\newcommand{\bee}{\begin{equation}}   
\newcommand{\ee}{\end{equation}}   
\newcommand{\beqn}{\begin{eqnarray}}   
\newcommand{\eeqn}{\end{eqnarray}}   
\newcommand{\fP}{I\!\!P}

\newcommand{\kf}{{\bf k}}   
\newcommand{\rf}{{\bf r}}   
\newcommand{\lf}{{\bf l}}   
   
\newcommand{\ef}{{\bf e}}   
\newcounter{savefig}

{\end{list}}   
\newcounter{enumct}

\newcommand{\captive}[1]{\rule{10mm}{0mm}%
\begin{minipage}{150mm}\caption[small]{#1}\end{minipage}}   
\begin{document}   
\noindent
DESY 99-027                           \hfill ISSN 0418--9833 \\
DTP/99/10\\
LUNFD6/(NFFL-7166) 1999  \\
\begin{center}   
\begin{Large}   
\vspace*{0.5cm}  
{\bf Quark-Antiquark-Gluon Jets in DIS Diffractive Dissociation}\\   
\end{Large}   
\vspace{0.5cm}   
J. Bartels$^a$ \footnote   
        {Supported by the TMR Network "QCD and Deep Structure of Elementary 
     Particles"}  
H. Jung$^b$ and M. W\"usthoff$^c$\hspace*{0.1cm}$^1$\\   
\vspace{0.5cm}   
$^a$ II. Inst. f. Theoretische Physik,    
Univ. Hamburg, Luruper Chaussee 149, D-22761 Hamburg\\   
$^b$ Department of Physics, Lund University, 221 00 Lund, Sweden\\   
$^c$  Department of Physics, University of Durham, Durham DH1 3LE, UK  
\end{center}   
\vspace*{0.75cm}   
   
\begin{abstract}   
We study the diffractive production of $q\bar{q}g$ jets with large transverse   
momenta in the region of large diffractive masses (small $\beta$).   
Cross sections for transverse and longitudinal photons are obtained in the   
leading log $1/x_{\fP}$ and log $1/\beta$ approximation, keeping all powers   
in log $k_t^2/Q^2$. We perform a numerical study and illustrate the   
angular distribution of the three jets. We also estimate the integrated   
diffractive three jet cross section and compare with the dijet cross section   
obtained before.   
\end{abstract}   
   
\noindent   
\section{Introduction}   
\setcounter{equation}{0}   
The diffractive dissociation in deep    
inelastic scattering lies on the interface between perturbative and    
non-perturbative QCD, and one expects the diffractive cross section to contain    
both hard and soft contributions. As a first experimental hint in this    
direction, one might take the observation that the energy dependence of   
the diffractive cross section, i.e. the measured intercept of the    
Pomeron flux factor tends to be    
larger than that of the pure soft Pomeron~\cite{H1,ZEUS}. As a possible   
interpretation of this observation, the contribution of the ``perturbative   
Pomeron``, the gluon structure function, may turn out to be substantial:    
the observed rise of the gluon structure function at small - $x$ enhances the   
hard part of the diffractive final state and thus leads to an effective   
Pomeron intercept which is bigger than that of the soft Pomeron. 
\par   
In a first attempt to provide a description of the diffractive cross section  
which takes into account both hard and soft contributions a parameterization  
has been suggested ~\cite{BEKW} and successfully fitted to both   
ZEUS and H1 ~\cite{BEKW,BR} data. As a main result, it was shown that   
an acceptable description of the data requires (at least) three pieces: the diffractive   
production of transverse $q\bar{q}$ and $q\bar{q}g$ states and the   
diffractive production of longitudinal vector mesons. The fits also confirmed   
that the effective intercept of the Pomeron tends to be larger than the  
soft Pomeron. 
\par  
In order to obtain further insight into DIS diffraction and, in particular,  
to understand the origin of the higher Pomeron intercept, one has to open   
the diffractive   
final state and to isolate those parts  
which can be attributed to the hard    
Pomeron. Apart from the longitudinal vector particles which have been suggested   
already several years ago~\cite{vector}, diffractive final states    
consisting of hard jets only,    
are the most promising candidates. As a first step, the production of    
quark-antiquark jets (Fig.~\ref{dijet}) with large transverse momenta has been    
discussed by several groups~\cite{BLW,NZ,LGM}.   
To summarize the main result, the energy dependence is given    
by the square of the gluon structure function:   
\beqn   
d\sigma \sim [\frac{1}{k_T^2}x_{\fP}g(x_{\fP}, k_T^2/(1-\beta))]^2   
\eeqn   
where $k_T$ denotes the transverse momentum of the quarks in the   
photon - Pomeron  
rest frame, and $\beta=Q^2 / (Q^2+M^2)$ is related to the mass $M$   
of the diffractive system. From eq.~(1.1) one concludes that for small masses    
($\beta$ near 1) the cross section will be dominated by this ``hard Pomeron``,   
either in the production of vector particles or of hard $q\bar{q}$-jets.   
For larger diffractive masses the contribution of hard $q\bar{q}$ jets is   
smaller but not negligible.  
Another interesting feature is the azimuthal dependence of the   
$q\bar{q}$ cross section. It was found~\cite{BELW} that, in the photon -   
Pomeron rest frame, the two jets prefer to   
lie perpendicular to the scattering plane (defined by the photon - Pomeron   
direction and the outgoing electron momentum). This has to be compared with   
the photon - gluon fusion mechanism in the usual DIS final state, where the   
two outgoing quarks prefer to lie inside the scattering plane. The experimental   
verification of these theoretical predictions turns out to be rather difficult:   
to make the calculation applicable, the jet final state is not allowed to   
contain any soft Pomeron remnants; the presence of such pieces changes the   
situation rather radically. To exclude such a soft remnant requires   
high statistics of the data and a rather good resolution in the jet algorithm.   
First results have been reported in~\cite{H1jet}. 
\par  
\begin{figure}[htb]   
\begin{center}   
\begin{picture}(0,0)%
\includegraphics{paper62.fig1.pstex}%
\end{picture}%
\setlength{\unitlength}{2368sp}%
\begingroup\makeatletter\ifx\SetFigFont\undefined  
\def\x#1#2#3#4#5#6#7\relax{\def\x{#1#2#3#4#5#6}}%
\expandafter\x\fmtname xxxxxx\relax \def\y{splain}%
\ifx\x\y   
\gdef\SetFigFont#1#2#3{%
  \ifnum #1<17\tiny\else \ifnum #1<20\small\else  
  \ifnum #1<24\normalsize\else \ifnum #1<29\large\else  
  \ifnum #1<34\Large\else \ifnum #1<41\LARGE\else  
     \huge\fi\fi\fi\fi\fi\fi  
  \csname #3\endcsname}%
\else  
\gdef\SetFigFont#1#2#3{\begingroup  
  \count@#1\relax \ifnum 25<\count@\count@25\fi  
  \def\x{\endgroup\@setsize\SetFigFont{#2pt}}%
  \expandafter\x  
    \csname \romannumeral\the\count@ pt\expandafter\endcsname  
    \csname @\romannumeral\the\count@ pt\endcsname  
  \csname #3\endcsname}%
\fi  
\fi\endgroup  
\begin{picture}(5197,3154)(2341,-6149)  
\put(2341,-3211){\makebox(0,0)[lb]{\smash{\SetFigFont{12}{14.4}{rm}$\gamma^*$}}}  
\put(3376,-4876){\makebox(0,0)[lb]{\smash{\SetFigFont{20}{24.0}{rm}$\sum$}}}  
\end{picture}  
\end{center}   
\captive{Dijet Production. The exchanged gluons couple to the quarks in all possible   
ways.   
\label{dijet}}   
\end{figure}   
\begin{figure}[htb]   
\begin{center}   
\begin{picture}(0,0)%
\includegraphics{paper62.fig2.pstex}%
\end{picture}%
\setlength{\unitlength}{2368sp}%
\begingroup\makeatletter\ifx\SetFigFont\undefined  
\def\x#1#2#3#4#5#6#7\relax{\def\x{#1#2#3#4#5#6}}%
\expandafter\x\fmtname xxxxxx\relax \def\y{splain}%
\ifx\x\y   
\gdef\SetFigFont#1#2#3{%
  \ifnum #1<17\tiny\else \ifnum #1<20\small\else  
  \ifnum #1<24\normalsize\else \ifnum #1<29\large\else  
  \ifnum #1<34\Large\else \ifnum #1<41\LARGE\else  
     \huge\fi\fi\fi\fi\fi\fi  
  \csname #3\endcsname}%
\else  
\gdef\SetFigFont#1#2#3{\begingroup  
  \count@#1\relax \ifnum 25<\count@\count@25\fi  
  \def\x{\endgroup\@setsize\SetFigFont{#2pt}}%
  \expandafter\x  
    \csname \romannumeral\the\count@ pt\expandafter\endcsname  
    \csname @\romannumeral\the\count@ pt\endcsname  
  \csname #3\endcsname}%
\fi  
\fi\endgroup  
\begin{picture}(6093,3675)(2040,-3892)  
\put(2040,-504){\makebox(0,0)[lb]{\smash{\SetFigFont{11}{13.2}{rm}$\gamma^*$}}}  
\put(3466,-2176){\makebox(0,0)[lb]{\smash{\SetFigFont{20}{24.0}{rm}$\sum$}}}  
\end{picture}  
\end{center}   
\captive{Three jet Production. The exchanged gluons couple to the quarks in all possible   
ways.   
\label{threejet}}   
\end{figure} 
For not so small diffractive masses ($\beta$ away from 1) one expects   
the production of extra gluons to become essential. The simplest final   
state consists of the $q\bar{q}$ pair and one gluon; in the region of large   
diffractive masses (small $\beta$) one expects a separation in rapidity   
between the $q\bar{q}$ system and the gluon (Fig.~\ref{threejet}) (not to be  
confused with the rapidity gap between the diffractive $q\bar{q}g$ system   
and the outgoing proton).  
In order to isolate    
the ``hard Pomeron`` contributions one, again, wants to study final states    
which consist only of    
jets with large transverse momenta (and no further soft remnants).   
In this paper we present the calculation of this three-jet cross section,   
restricting ourselves to the region of large diffractive masses (small   
$\beta$). We shall work in   
the leading - $\log(1/\beta$), $\log(1/x_{\fP}$) approximation, retaining all  
powers in $k_T^2/Q^2$. This approximation includes also higher-twist   
terms of the diffractive cross section. Diffractive production of $q\bar{q}g$   
final states in another kinematic region has been studied before~\cite{R,GBW}:   
the gluon transverse momentum has to be smaller than those of the quarks.   
The results of~\cite{GBW} extend into the large $\beta$-region.    
\par 
There are several questions to which our jet analysis gives an answer.  
Apart from a quantitative estimate of the jet rates (as a function of the  
imposed kinematical cuts) our analytic formula allows a first step towards an   
analysis of the event shape of diffractive final states (e.g. thrust).   
In particular, we  
find that in the region of medium small $\beta$ (where our formula is   
applicable) the $q\bar{q}g$ final state provides a substantial contribution  
to the two-jet cross section. Our framework also allows to address the  
transition from hard jets to the aligned gluon configuration: in a future step  
one might model a two gluon amplitude which allows to extrapolate our  
formula into the soft Pomeron region.   
Such a model (an explicit parameterization was suggested   
in ref.~\cite{GBW2}) would allow to  
describe both the hard and the soft region of diffractive $q\bar{q}g$  
production. 
However, a severe limitation is our restriction to the small-$\beta$   
region. In order to cover the full $\beta$ range we need a complete  
LO calculation of three parton final states in DIS diffraction which includes  
a NLO two jet calculation.       
\par 
This paper will be organized as follows. In section 2 we describe our   
calculation and present the result for the three jet production cross  
section. Section 3 contains the transformation into the impact parameter space  
and in section 4 we discuss a few particular kinematic regions   
for which our formula allows analytic approximations.   
In section 5 we perform a first numerical   
analysis, calculate integrated cross sections and show a few plots. In   
the final section we summarize our results and make a few   
concluding remarks.     
\section{The Differential Jet Production Cross Section}  
\setcounter{equation}{0}   
Our notations are illustrated in Fig.~\ref{notation}.  
 As usual, $q$ and $p$ denote   
the four momenta of the photon and the proton,  
 respectively, $t=r^2$ is the square    
of the momentum transferred from the proton to the diffractive system.   
The total energy is denoted by $s$ and the sub-energy    
of the $\gamma^*$-proton system with $W^2$.   
\begin{figure}[htb]   
\begin{center}   
\begin{picture}(0,0)%
\includegraphics{paper62.fig3.pstex}%
\end{picture}%
\setlength{\unitlength}{2368sp}%
\begingroup\makeatletter\ifx\SetFigFont\undefined  
\def\x#1#2#3#4#5#6#7\relax{\def\x{#1#2#3#4#5#6}}%
\expandafter\x\fmtname xxxxxx\relax \def\y{splain}%
\ifx\x\y   
\gdef\SetFigFont#1#2#3{%
  \ifnum #1<17\tiny\else \ifnum #1<20\small\else  
  \ifnum #1<24\normalsize\else \ifnum #1<29\large\else  
  \ifnum #1<34\Large\else \ifnum #1<41\LARGE\else  
     \huge\fi\fi\fi\fi\fi\fi  
  \csname #3\endcsname}%
\else  
\gdef\SetFigFont#1#2#3{\begingroup  
  \count@#1\relax \ifnum 25<\count@\count@25\fi  
  \def\x{\endgroup\@setsize\SetFigFont{#2pt}}%
  \expandafter\x  
    \csname \romannumeral\the\count@ pt\expandafter\endcsname  
    \csname @\romannumeral\the\count@ pt\endcsname  
  \csname #3\endcsname}%
\fi  
\fi\endgroup  
\begin{picture}(4551,2505)(2005,-3518)  
\put(6532,-1758){\makebox(0,0)[lb]{\smash{\SetFigFont{10}{12.0}{rm}$k_1$}}}  
\put(6544,-2272){\makebox(0,0)[lb]{\smash{\SetFigFont{10}{12.0}{rm}$k_2$}}}  
\put(2041,-1221){\makebox(0,0)[lb]{\smash{\SetFigFont{10}{12.0}{rm}q}}}  
\put(6521,-1185){\makebox(0,0)[lb]{\smash{\SetFigFont{10}{12.0}{rm}$r+q-k_1-k_2$}}}  
\put(2005,-3347){\makebox(0,0)[lb]{\smash{\SetFigFont{10}{12.0}{rm}p}}}  
\put(6556,-3323){\makebox(0,0)[lb]{\smash{\SetFigFont{10}{12.0}{rm}p'=p-r}}}  
\end{picture}  
\end{center}   
\captive{Notations for the diffractive production of a $q\bar{q}g$   
system.   
\label{notation}}   
\end{figure}   
The scaling variables are $x=Q^2/2pq$, $x_{\fP}=(Q^2+M^2)/(Q^2+W^2)$ , and   
$\beta=Q^2/(Q^2+M^2)$. For simplicity we restrict ourselves to the forward   
direction, $t=0$, such that the four momentum $r=x_{\fP}p$ has no transverse    
component. We use the Sudakov parameterization   
\beqn   
k_i = \alpha_i q' + \beta_i p +k_{i\;T},\;\; \kf_i^2 = - k_{i\;T}^2,   
\eeqn   
where $q'=q + xp$. We work in a reference frame where photon and proton   
momenta are collinear (along the $z$ - direction); the scattering plane (the   
$x-z$ plane) is defined by the incoming and outgoing electron. Azimuthal angles   
refer to this scattering plane. In order to justify the use of perturbation   
theory we start in the ``hard region``: all three final state partons have   
large transverse momenta. Later on we will show that the more precise   
conditions imply that the virtualities of the exchanged   
quark and gluon have to be large. This leads to:   
\beqn   
Q_0^2 &<& \kf_2^2 \nonumber \\   
Q_0^2 &<& \frac{1}{\alpha_1}[\alpha_1(1-\alpha_1) Q^2 +\kf_1^2]   
\nonumber \\   
Q_0^2 &<& \frac{1}{1-\alpha_1}[\alpha_1(1-\alpha_1) Q^2 +(\kf_1+\kf_2)^2]   
\eeqn   
Our calculation will be done in the   
leading - $\log W^2$, leading - $\log M^2$ approximation and applies   
to the region $Q^2 \ll M^2 \ll W^2$. As usual, in this approximation   
we have    
$\alpha_2 \ll \alpha_1, 1-\alpha_1$ and $\beta_2$ close to $x_{\fP}$.   
For large values of   
$\log Q^2/\kf _i^2$ our approximation reduces to the double logarithmic    
approximation (DLA).   
The invariant mass $M$ of the diffractive system can be expressed (in 
this leading log approximation) as   
\beqn   
M^2 = W^2 (x_{\fP} - x)\;\;.  
\eeqn   
The mass of the $q\bar{q}$ subsystem is given by  
$m_{q\bar{q}}^2=m^2-\kf_2^2$ with  
\beqn   
m^2=M^2 - \beta_2 W^2, \;\; \beta_2 =\frac{M^2-m^2}{W^2}\;\;.  
\eeqn   
For $Q^2\ll M^2$ (i.e. small $\beta$) we expect the invariant mass of the   
$q\bar{q}$ pair to be much smaller than $M^2$ (of the order of $Q^2$).   
From the mass shell condition of the upper quark line (with momentum   
$r+q-k_1-k_2$) we have the relation:   
\beqn   
0&=&\alpha_1 (1-\alpha_1-\alpha_2) (M^2-\beta_2 W^2) -(1-\alpha_1-\alpha_2)    
\kf_1^2   
- \alpha_1 (\kf_1 + \kf_2)^2    
\nonumber  \\   
&\approx &\alpha_1 (1-\alpha_1) m^2 -(1-\alpha_1) \kf_1^2   
- \alpha_1 (\kf_1 + \kf_2)^2   
\eeqn    
We begin with the calculation of the $\gamma^*p$ cross   
section. The diagrams we have to calculate are shown in Fig.~\ref{diag}.    
Since we consider the square of the scattering    
matrix element and, moreover, take the discontinuity in $W^2$, all quark lines    
between the leftmost and the rightmost   
gluon are on-shell; the same holds for the produced gluon below the   
quark loop. This leads to $\delta$-functions which will be used to   
integrate over all $\alpha_i$ and $\beta_i$, except for $\beta_2$.   
We begin with the transverse cross section.   
Let $e_x=(0,1,0,0)=(0,\ef_x,0)$ and $e_y=(0,0,1,0)=(0,\ef_y,0)$    
be the transverse polarization vectors    
of the photon (it should be noted that $e_x$ lies in the scattering plane).   
Contracting the Lorentz indices of the two external photons    
with the transverse polarization vectors $e_1$ and $e_2$ (each of them    
can be either $e_x$ or $e_y$), the sum of the diagrams in Fig.~\ref{diag}  
 leads to a    
contribution of the form   
\beqn   
2 W^4 \delta \left( (r+q-k_1-k_2)^2 \right)   
\left[4 \alpha_1^2(1-\alpha_1)^2  \ef_{1i} \ef_{2j} M_{il} M'_{jl}   
-\alpha_1(1-\alpha_1) \ef_1 \cdot \ef_2 M_{il}M'_{il} \right]   
\eeqn   
(here $i,j,l=1,2$  refer to the transverse components; repeated indices are    
summed over, and the prime at the   
second $M'_{il}$ factor indicates that it belongs to the   
two-gluon amplitude with momentum $\lf'$, attached to the lower two gluons on  
the rhs of the diagrams in Fig.4).  
 We have exhibited the   
$\delta$ function of the upper quark line which leads to the condition   
eq.~(2.4) and will be used to   
express $\alpha_1$ in terms of $M$, $\beta_2$, $\kf_1$, and $\kf_2$. The   
expression for $M_{ij}$ will be discussed below. Contracting this with the   
lepton tensor $\frac{y^2}{4Q^2} L_{\mu \nu} e_1^{\mu} e_2^{\nu}$ and summing    
over the two possible transverse polarizations we obtain    
\beqn   
2 W^4 \delta \left( (r+q-k_1-k_2)^2 \right) \hspace{7cm}   
\nonumber \\\cdot \left[ \frac{1+(1-y)^2}{2} 4 \alpha_1(1-\alpha_1)   
             [\alpha_1^2 + (1-\alpha_1)^2]   
M_{il}M'_{il} - 4 (1-y) \alpha_1^2(1-\alpha_1)^2    
(M_{1l}M'_{1l} - M_{2l}M'_{2l}) \right]   
\eeqn   
In the second line one recognizes two terms. The first one    
(proportional to $M_{il}M'_{il}$) corresponds to the sum    
over the two transverse polarizations, the second one    
(proportional to $M_{1l}M'_{1l} - M_{2l}M'_{2l}$, $M_{1l}=\ef_{xi}M_{il}$,   
$M_{2l}=\ef_{yi}M_{il}$)    
to the difference.   
As long as we do not integrate over the azimuthal angle of the   
outgoing electron, the two transverse polarizations of the photon come with   
different weights, and that is why we write our result in terms of   
both the sum and the difference of the two polarizations. 
\par  
\begin{figure}   
\begin{center}   
\begin{picture}(0,0)%
\includegraphics{paper62.fig4.pstex}%
\end{picture}%
\setlength{\unitlength}{3947sp}%
\begingroup\makeatletter\ifx\SetFigFont\undefined  
\def\x#1#2#3#4#5#6#7\relax{\def\x{#1#2#3#4#5#6}}%
\expandafter\x\fmtname xxxxxx\relax \def\y{splain}%
\ifx\x\y   
\gdef\SetFigFont#1#2#3{%
  \ifnum #1<17\tiny\else \ifnum #1<20\small\else  
  \ifnum #1<24\normalsize\else \ifnum #1<29\large\else  
  \ifnum #1<34\Large\else \ifnum #1<41\LARGE\else  
     \huge\fi\fi\fi\fi\fi\fi  
  \csname #3\endcsname}%
\else  
\gdef\SetFigFont#1#2#3{\begingroup  
  \count@#1\relax \ifnum 25<\count@\count@25\fi  
  \def\x{\endgroup\@setsize\SetFigFont{#2pt}}%
  \expandafter\x  
    \csname \romannumeral\the\count@ pt\expandafter\endcsname  
    \csname @\romannumeral\the\count@ pt\endcsname  
  \csname #3\endcsname}%
\fi  
\fi\endgroup  
\begin{picture}(4355,8527)(2951,-8475)  
\put(5311,-241){\makebox(0,0)[lb]{\smash{\SetFigFont{9}{10.8}{rm}$k_1$}}}  
\put(5437,-1644){\makebox(0,0)[lb]{\smash{\SetFigFont{9}{10.8}{rm}$l'$}}}  
\put(5653,-1644){\makebox(0,0)[lb]{\smash{\SetFigFont{9}{10.8}{rm}$l'+r$}}}  
\put(4070,-44){\makebox(0,0)[lb]{\smash{\SetFigFont{9}{10.8}{rm}$q$}}}  
\put(5223,-1949){\makebox(0,0)[lb]{\smash{\SetFigFont{10}{12.0}{rm}a}}}  
\put(4677,-1644){\makebox(0,0)[lb]{\smash{\SetFigFont{9}{10.8}{rm}$l+r$}}}  
\put(5311,-826){\makebox(0,0)[lb]{\smash{\SetFigFont{9}{10.8}{rm}$k_2$}}}  
\put(5223,-8475){\makebox(0,0)[lb]{\smash{\SetFigFont{10}{12.0}{rm}c}}}  
\put(5223,-5350){\makebox(0,0)[lb]{\smash{\SetFigFont{10}{12.0}{rm}b}}}  
\put(5060,-1642){\makebox(0,0)[lb]{\smash{\SetFigFont{9}{10.8}{rm}$l$}}}  
\end{picture}  
\end{center}   
\captive{Diagrams for the cross section of diffractive $q\bar{q}g$    
production. We only show the diffractive system: at the lower end the    
two-gluon systems   
on both sides couple to the gluon structure function of the proton.   
At the upper end, the gluons couple to the quark loop in all possible ways.  
We do not show the vertical discontinuity line which divides each diagram   
into left and right hand pieces.    
\label{diag}}   
\end{figure}  
For the longitudinal cross section we arrive at the following structure:   
\beqn   
2 W^4 \delta \left( (r+q-k_1-k_2)^2 \right)   
\alpha_1^3 (1-\alpha_1)^3 Q^2 M_l M'_l.   
\eeqn   
Together with the lepton tensor we obtain   
\beqn   
2 W^4 \delta \left( (r+q-k_1-k_2)^2 \right)   
2 (1-y) \alpha_1^3 (1-\alpha_1)^3 Q^2 M_l M'_l.   
\eeqn   
\par  
Finally, we have the interference between the transverse and the longitudinal    
photon. The expressions analogous to eq.~(2.5) and eq.~(2.6) are given by   
\beqn   
2 W^4 \delta \left( (r+q-k_1-k_2)^2 \right)    
2\sqrt{Q^2} \alpha_1^2 (1-\alpha_1)^2 (1-2\alpha_1)    
M_{1l} M'_l.   
\eeqn   
and    
\beqn   
2 W^4 \delta \left( (r+q-k_1-k_2)^2 \right)    
 \sqrt{1-y} (2-y) \sqrt{Q^2} \alpha_1^2 (1-\alpha_1)^2 (1-2\alpha_1)    
M_{1l} M'_l.   
\eeqn   
\par   
Before we turn to the calculation of the $M_{il}$ and $M_l$, we write down    
the final formula for the process   
$e + p \to e' + q\bar{q}g + p$. The diffractive final state is characterized   
by the invariant mass $M$, the longitudinal momentum fraction $\beta_2$ of the   
outgoing gluon, and by the transverse momenta $\kf_1$, $\kf_2$ of the    
outgoing quark and gluon, respectively    
(in our notation, $\kf^2=-k_T^2>0$). Starting from eq.~(2.5) - eq.~(2.10) and   
evaluating the final state phase space integrals we obtain:   
\beqn   
\frac{ d \sigma^{e^- p}_D}   
{d y d Q^2 d M^2 d m^2 d^2 \kf_1 d^2 \kf_2  d t}_{|t=0}   
= \frac{\alpha_{em}}{ y Q^2 \pi} \hspace{4cm} \nonumber \\   
\cdot \left[   
\frac{1+(1-y)^2}{2}   
\frac{d \sigma^{\gamma^* p}_{D,T,+}}  
{d M^2 d m^2 d^2 \kf_1 d^2 \kf_2 d t}_{|t=0}   
- 2 (1-y)     
\frac{d \sigma^{\gamma^* p}_{D,T,-}}  
{d M^2 d m^2 d^2 \kf_1 d^2 \kf_2 d t}_{|t=0}\right.   
\nonumber \\    
\left. + (1-y) \frac{d \sigma^{\gamma^* p}_{D,L}}   
           {d M^2 d m^2 d\kf_1 d^2 \kf_2 d t}_{|t=0}   
+(2-y)\sqrt{1-y}    
\frac{d \sigma^{\gamma^* p}_{D,I}}{d M^2 d m^2 d^2 \kf_1 d^2 \kf_2 d t}_{|t=0}   
\right], \;\;\;\;\;\;   
\label{eq1}   
\eeqn   
where the $\gamma^*p$ cross sections are defined as:   
\beqn   
\frac{d \sigma^{\gamma^* p}_{D,T+}}{d M^2 d m^2 d^2 \kf_1 d^2 \kf_2   
d t}_{|t=0}   
&=&\frac{9}{128\pi}\frac{1}{\sqrt{S} (M^2-m^2) m^2}   
          \sum_f e_f^2 \alpha_{em} \alpha_s^3   
\left[\alpha_1^2 + (1-\alpha_1)^2\right] \alpha_1 (1-\alpha_1) \nonumber \\   
&& \hspace{2cm} \cdot M_{il} M'_{il}\\   
\frac{d \sigma^{\gamma^* p}_{D,T-}}{d M^2 d m^2 d^2 \kf_1 d^2 \kf_2    
d t}_{|t=0}   
&=&\frac{9}{128\pi}\frac{1}{\sqrt{S} (M^2-m^2) m^2}   
            \sum_f e_f^2 \alpha_{em} \alpha_s^3   
\alpha_1^2 (1-\alpha_1)^2 \cdot \nonumber \\   
&& \hspace{2cm} \left(M_{1l} M'_{1l} - M_{2l} M'_{2l}\right)\\   
\frac{d \sigma^{\gamma^* p}_{D,L}}{d M^2 d m^2 d^2 \kf_1 d^2 \kf_2    
d t}_{|t=0}   
&=&\frac{9}{128\pi}\frac{1}{\sqrt{S} (M^2-m^2) m^2}   
         \sum_f e_f^2 \alpha_{em} \alpha_s^3   
4\alpha_1^3 (1-\alpha_1)^3 Q^2 M_l M'_l\\   
\frac{d \sigma^{\gamma^* p}_{D,I}}{d M^2 d m^2 d^2 \kf_1 d^2 \kf_2    
d t}_{|t=0}   
&=&\frac{9}{128\pi}\frac{1}{\sqrt{S} (M^2-m^2) m^2}   
         \sum_f e_f^2 \alpha_{em} \alpha_s^3   
\alpha_1^2 (1-\alpha_1)^2 (1-2\alpha_1) \nonumber \\   
&& \hspace{2cm} \cdot   
        \frac{1}{2}\sqrt{Q^2} \left[M_{1l}M'_l + M_l M'_{1l}\right].   
\eeqn   
Here   
\beqn   
S=\left(1+\frac{\kf_1^2}{m^2}-\frac{(\kf_1+\kf_2)^2}{m^2}   
\right)^2 - 4\frac{\kf_1^2}{m^2},   
\eeqn   
and the prefactors containing $\alpha_1$ have to be re-expressed using    
the $\delta$-function condition eq.~(2.4). As we have    
discussed before, the transverse cross section has two terms.  
 The first one corresponds to the sum    
over the two transverse polarizations, the second one to the difference.   
This second term of the transverse cross section, as well as the    
interference cross section, introduce the     
azimuthal dependence of our cross section. 
\par  
\begin{figure}[htb]   
\begin{center}   
\begin{picture}(0,0)%
\includegraphics{paper62.fig5.pstex}%
\end{picture}%
\setlength{\unitlength}{3158sp}%
\begingroup\makeatletter\ifx\SetFigFont\undefined  
\def\x#1#2#3#4#5#6#7\relax{\def\x{#1#2#3#4#5#6}}%
\expandafter\x\fmtname xxxxxx\relax \def\y{splain}%
\ifx\x\y   
\gdef\SetFigFont#1#2#3{%
  \ifnum #1<17\tiny\else \ifnum #1<20\small\else  
  \ifnum #1<24\normalsize\else \ifnum #1<29\large\else  
  \ifnum #1<34\Large\else \ifnum #1<41\LARGE\else  
     \huge\fi\fi\fi\fi\fi\fi  
  \csname #3\endcsname}%
\else  
\gdef\SetFigFont#1#2#3{\begingroup  
  \count@#1\relax \ifnum 25<\count@\count@25\fi  
  \def\x{\endgroup\@setsize\SetFigFont{#2pt}}%
  \expandafter\x  
    \csname \romannumeral\the\count@ pt\expandafter\endcsname  
    \csname @\romannumeral\the\count@ pt\endcsname  
  \csname #3\endcsname}%
\fi  
\fi\endgroup  
\begin{picture}(5312,4568)(2086,-5277)  
\put(2829,-1555){\makebox(0,0)[lb]{\smash{\SetFigFont{11}{13.2}{rm}1}}}  
\put(6409,-5277){\makebox(0,0)[lb]{\smash{\SetFigFont{11}{13.2}{rm}1}}}  
\put(7254,-5267){\makebox(0,0)[lb]{\smash{\SetFigFont{11}{13.2}{rm}$k_1^2/m^2$}}}  
\put(1500,-985){\makebox(0,0)[lb]{\smash{\SetFigFont{11}{13.2}{rm}$(k_1+k_2)^2/m^2$}}}  
\put(3459,-4362){\makebox(0,0)[lb]{\smash{\SetFigFont{11}{13.2}{rm}$S>0$}}}  
\end{picture}  
\end{center}   
\captive{Kinematic Boundaries. The drawn line belongs to $S=0$.   
\label{Kin}}   
\end{figure}   
It is instructive to consider the kinematics of the three partons in the final    
state. Within our approximation, we have to restrict ourselves to the   
forward direction of the Pomeron: in the $\gamma^*$-$\fP$ CM system the gluon   
is emitted in the direction of the Pomeron.   
The kinematic boundaries of the transverse   
momenta $\kf_1$ and $\kf_2$ are determined by the condition $S>0$ (eq.~(2.17)).   
The allowed region is illustrated in Fig.~\ref{Kin},  
 where we have plotted the two   
quark transverse momenta, normalized to the invariant mass square $m^2$ of the    
$q\bar{q}$ subsystem: $\kf_1^2/m^2$ and $(\kf_1+\kf_2)^2/m^2$. The curve   
$S=0$ shows that the quark momenta are essentially restricted by $m^2$.   
For the particular case of   
small gluon transverse momenta $\kf_2^2 \ll \kf_1^2$, we are on the   
diagonal line $\kf_1^2 \approx (\kf_1+\kf_2)^2$, and the physical region is   
restricted to $\kf_1^2 \le \frac{1}{4} m^2$. For a given set of transverse   
momenta, the longitudinal momentum fraction $\alpha_1$ is given by one of     
the two values:   
\be   
\alpha_1=\frac{1}{2}\left( 1+\frac{\kf_1^2 - (\kf_1+\kf_2)^2}{m^2}   
 \pm \sqrt{S} \right),   
\ee   
($\alpha_2$, $\beta_1$, and $\beta_2$ follow from the mass shell conditions   
$k_1^2=0$, $k_2^2=0$). The limiting values $\alpha_1=0$ and $1-\alpha_1=0$ lie    
on the axis $\kf_1^2=0$ and $(\kf_1+\kf_2)^2=0$, respectively,  
 i.e. they correspond    
to final states where one of the quark lines becomes soft.   
For small $\kf_2^2$ (i.e. on the line $\kf_1^2 = (\kf_1+\kf_2)^2$)   
eq.~(2.17) simplifies to    
\beqn   
\alpha_1=\frac{1}{2}(1\pm \sqrt{1-4\frac{\kf_1^2}{m^2}})   
\eeqn   
and    
\beqn   
\alpha_1 (1-\alpha_1)=\kf_1^2/m^2.   
\eeqn   
For the angle $\Delta$ between the two quarks we find   
\be   
\frac{1-\cos \Delta}{2} =    
\frac{ (\kf_1 +\alpha_1 \kf_2)^2 / M^2}   
     {[\alpha_1^2 + \kf_1^2/M^2][(1-\alpha_1)^2+(\kf_1+\kf_2)^2/M^2]}   
\ee   
In the denominator, the $\alpha_1$ terms will dominate, as long as we stay away    
from the values $\alpha_1=0,1$, the numerator is (at  most) of the   
order $m^2/M^2$ and the opening angle $\Delta=O(m^2/M^2)$. Therefore, for   
small $m^2/M^2$ the opening angle will be small (of the order    
$\Delta=O(m^2/M^2)$),   
and we expect, in the $\gamma^* - \PO$ CM system, the gluon and   
the $q\bar{q}$-pair to come out as two back-to-back jets. The $q\bar{q}$ - pair   
produces a broader jet, where the opening angle is given by eq.~(2.21) and will    
approximately be of the order of $O(m^2/M^2)$. When $m^2$ gets larger,    
the $q\bar{q}$ - pair jet becomes broader and eventually can be resolved into    
two separate jets.   
The angle of the axis of the two-jet configuration is given by the direction    
of the outgoing gluon. If $\theta_2$ denotes the angle between the gluon and   
the proton momentum, we find:   
\be   
\frac{1-\cos \theta_2}{2} = \frac{\kf_2^2/M^2}{\kf_2^2/M^2 + (1-m^2/M^2)^2}.    
\ee   
For $m^2 \ll M^2$ the gluon transverse momentum $\kf_2^2$ will be much    
smaller than $M^2$, the angle will be small, i.e.   
the outgoing gluon moves in the direction of the proton with an angle   
of the order  $O(\kf_2^2/M^2)$. In the strong ordering region    
$\kf_2^2 \ll \kf_1^2$, eq.~(2.21) can be    
approximated by   
\beqn   
\frac{1-\cos \Delta}{2}=\frac{m^2}{M^2} \frac{1}   
     {[\alpha_1+(1-\alpha_1)m^2/M^2][(1-\alpha_1)+\alpha_1 m^2/M^2]}   
\eeqn   
\par 
In section 5 we present results of a numerical study of our cross section   
formula which illustrate the geometry of the three-jet configuration.  
\par   
Let us now turn to the calculation of the $M_{il}$ and $M_l$.   
Each of them is of the form   
\beqn    
M_{il}&=& \int \frac{d^2\lf}{\pi \lf^2} {\cal F}(x_{\fP},\lf^2) T_{il},   
\eeqn   
(and an analogous expression for $M_l$), where ${\cal F}$ denotes the    
unintegrated gluon structure function   
\beqn   
\int_{Q_0^2}^{Q^2} d\lf^2 {\cal F}(x, \lf^2) = x g(x, Q^2)   
\eeqn   
and models the perturbative Pomeron.   
In the fermion loop at the top of the diagrams in Fig.~\ref{diag}   
it is understood that    
the gluons couple to the quarks in   
all possible ways. Since we consider the square of the scattering    
matrix element and take the discontinuity in $W^2$, all quark lines    
between the leftmost and the rightmost   
gluon are on - shell; this leads to $\delta$ - functions which will be used to   
integrate over all $\alpha_i$ and $\beta_i$, except for $\beta_2$.    
We begin with the transverse cross section.   
The diagrams with two gluons lines attached to the fermion loop   
(Fig.~\ref{diag}$a$) lead to:    
\beqn   
T_{il}^{(a)} &=&2\;\;   
\left( \frac{\kf_1+\kf_2}{D(\kf_1 + \kf_2)} -    
\frac{\kf_1}{D(\kf_1)} \right)_i   
\left(\frac{\kf_2 +\lf}{(\kf_2+\lf)^2} - \frac{\kf_2}{\kf_2^2} \right)_l,   
\eeqn   
with   
\beqn   
D(\kf)=\alpha_1(1-\alpha_1)Q^2 + \kf^2.   
\eeqn   
In eq.~(2.26) the first factor (with subscript `$i$`)   
comes from the fermion loop, the second factor from   
gluon production below the fermion loop. For the latter we have made use of    
the $K_{2 \to 4}$ gluon vertex~\cite{B,BW}:   
\beqn   
\left(\frac{\kf_2 +\lf}{(\kf_2+\lf)^2} - \frac{\kf_2}{\kf_2^2}  \right)_l   
\left(\frac{\kf_2 +\lf'}{(\kf_2+\lf')^2} - \frac{\kf_2}{\kf_2^2} \right)_l   
\eeqn   
(this includes the two vertical gluon propagators with momenta $\kf_2$).   
A similar expression holds for $M'_{jl}$.   
\par   
Next we turn to the four diagrams in Fig.~\ref{diag}$b$ where four gluons are    
attached to the fermion loop. We write :   
\beqn   
(M_{il} M'_{jl})^{(b1)}=\int \frac{d^2\lf}{\pi \lf^2} {\cal F}(x_{\fP},\lf^2)    
\int \frac{d^2\lf'}{\pi \lf'^2} {\cal F}(x_{\fP},{\lf^{\prime}}^2)    
                         (T_{il} T'_{jl})^{(b1)}   
\eeqn   
The first diagram in Fig.~\ref{diag}$b$ leads to:   
\beqn   
(T_{il} T'_{jl})^{(b1)}   
      =\left(\frac{\lf + \kf_1 + \kf_2}{D(\lf+\kf_1 + \kf_2)}   
     + \frac{\lf - \kf_1}{D(\lf - \kf_1)}   
     - \frac{\kf_1 + \kf_2}{D(\kf_1 + \kf_2)}   
     +\frac{\kf_1}{D(\kf_1)} \right)_i    
      \left( \frac{\kf_2 +\lf}{(\kf_2+\lf)^2}- \frac{\kf_2}{\kf_2^2} \right)_l   
\nonumber \\   
\cdot       \left(\frac{\lf' + \kf_1 + \kf_2}{D(\lf' +\kf_1 + \kf_2)}   
     + \frac{\lf' - \kf_1}{D(\lf' - \kf_1)}   
     - \frac{\kf_1 + \kf_2}{D(\kf_1 + \kf_2)}   
     +\frac{\kf_1}{D(\kf_1)} \right)_j   
      \left( \frac{\kf_2 +\lf'}{(\kf_2+\lf')^2}- \frac{\kf_2}{\kf_2^2}\right)_l   
\eeqn   
Similarly, we find for the other three contributions:   
\beqn   
(T_{il} T'_{jl})^{(b2)}   
      =-\left(\frac{\lf - \kf_1 - \kf_2}{D(\lf - \kf_1 - \kf_2)}   
     + \frac{\lf + \kf_1}{D(\lf + \kf_1)}   
     - \frac{\kf_1 + \kf_2}{D(\kf_1 + \kf_2)}   
     +\frac{\kf_1}{D(\kf_1)} \right)_i    
      \left( \frac{\kf_2 -\lf}{(\kf_2-\lf)^2}- \frac{\kf_2}{\kf_2^2}\right)_l   
\nonumber \\   
\cdot       \left(\frac{\lf' + \kf_1 + \kf_2}{D(\lf'+\kf_1 + \kf_2)}   
     + \frac{\lf' - \kf_1}{D(\lf' - \kf_1)}   
     - \frac{\kf_1 + \kf_2}{D(\kf_1 + \kf_2)}   
     +\frac{\kf_1}{D(\kf_1)} \right)_j   
      \left( \frac{\kf_2 +\lf'}{(\kf_2+\lf')^2}- \frac{\kf_2}{\kf_2^2}\right)_l,   
\eeqn   
\beqn   
(T_{il} T'_{jl})^{(b3)}   
      =-\left(\frac{\lf + \kf_1 + \kf_2}{D(\lf+ \kf_1 + \kf_2)}   
     + \frac{\lf - \kf_1}{D(\lf - \kf_1)}   
     - \frac{\kf_1 + \kf_2}{D(\kf_1 + \kf_2)}   
     +\frac{\kf_1}{D(\kf_1)} \right)_i    
      \left( \frac{\kf_2 +\lf}{(\kf_2+\lf)^2}- \frac{\kf_2}{\kf_2^2}\right)_l   
\nonumber \\   
\cdot       \left(\frac{\lf' - \kf_1 - \kf_2}{D(\lf' -\kf_1 - \kf_2)}   
     + \frac{\lf' + \kf_1}{D(\lf + \kf_1)}   
     - \frac{\kf_1 + \kf_2}{D(\kf_1 + \kf_2)}   
     +\frac{\kf_1}{D(\kf_1)} \right)_j   
      \left( \frac{\kf_2 -\lf'}{(\kf_2-\lf')^2}- \frac{\kf_2}{\kf_2^2}\right)_l,   
\eeqn   
and   
\beqn   
(T_{il} T'_{jl})^{(b4)}   
      =\left(\frac{\lf - \kf_1 - \kf_2}{D(\lf - \kf_1 - \kf_2)}   
     + \frac{\lf + \kf_1}{D(\lf + \kf_1)}   
     - \frac{\kf_1 + \kf_2}{D(\kf_1 + \kf_2)}   
     +\frac{\kf_1}{D(\kf_1)} \right)_i    
      \left( \frac{\kf_2 -\lf}{(\kf_2-\lf)^2}- \frac{\kf_2}{\kf_2^2}\right)_l   
\nonumber \\   
\cdot       \left(\frac{\lf' - \kf_1 - \kf_2}{D(\lf' - \kf_1 - \kf_2)}   
     + \frac{\lf' + \kf_1}{D(\lf' + \kf_1)}   
     - \frac{\kf_1 + \kf_2}{D(\kf_1 + \kf_2)}   
     +\frac{\kf_1}{D(\kf_1)} \right)_j   
      \left( \frac{\kf_2 -\lf'}{(\kf_2-\lf')^2}- \frac{\kf_2}{\kf_2^2}\right)_l   
\eeqn   
\par   
The third group in Fig.~\ref{diag}$c$   
consists of diagrams with three gluons being    
attached to the quark loop. We obtain:   
\beqn   
(T_{il} T'_{jl})^{(c1)}   
=2 \left( \frac{\kf_1+\kf_2}{D(\kf_1 + \kf_2)}    
                         -\frac{\kf_1}{D(\kf_1)} \right)_i    
\left(\frac{\kf_2 +\lf}{(\kf_2+\lf)^2} - \frac{\kf_2}{\kf_2^2} \right)_l   
\hspace{5cm}   
\nonumber \\   
\cdot\left(\frac{\lf' + \kf_1 + \kf_2}{D(\lf' +\kf_1 + \kf_2)}   
     + \frac{\lf' - \kf_1}{D(\lf' - \kf_1)}   
     - \frac{\kf_1 + \kf_2}{D(\kf_1 + \kf_2)}   
     +\frac{\kf_1}{D(\kf_1)} \right)_j   
      \left( \frac{\kf_2 +\lf'}{(\kf_2+\lf')^2}- \frac{\kf_2}{\kf_2^2}\right)_l   
\eeqn   
\beqn   
(T_{il} T'_{jl})^{(c2)}   
=-2 \left( \frac{\kf_1+\kf_2}{D(\kf_1 + \kf_2)} -    
                        \frac{\kf_1}{D(\kf_1)} \right)_i   
\left(\frac{\kf_2 +\lf}{(\kf_2+\lf)^2} - \frac{\kf_2}{\kf_2^2} \right)_l   
\hspace{5cm}    
\nonumber \\   
\cdot \left(\frac{\lf' - \kf_1 - \kf_2}{D(\lf' -\kf_1 - \kf_2)}   
     + \frac{\lf' + \kf_1}{D(\lf' + \kf_1)}   
     - \frac{\kf_1 + \kf_2}{D(\kf_1 + \kf_2)}   
     +\frac{\kf_1}{D(\kf_1)} \right)_j   
      \left( \frac{\kf_2 -\lf'}{(\kf_2-\lf')^2}- \frac{\kf_2}{\kf_2^2}\right)_l   
\eeqn   
\beqn   
(T_{il} T'_{jl})^{(c3)}   
=2    
\left(\frac{\lf + \kf_1 + \kf_2}{D(\lf+ \kf_1 + \kf_2)}   
     + \frac{\lf - \kf_1}{D(\lf - \kf_1)}   
     - \frac{\kf_1 + \kf_2}{D(\kf_1 + \kf_2)}   
     +\frac{\kf_1}{D(\kf_1)} \right)_i    
      \left( \frac{\kf_2 +\lf}{(\kf_2+\lf)^2}- \frac{\kf_2}{\kf_2^2}\right)_l   
\nonumber \\   
\cdot \left( \frac{\kf_1+\kf_2}{D(\kf_1 + \kf_2)} -    
                           \frac{\kf_1}{D(\kf_1)} \right)_j   
\left( \frac{\kf_2 +\lf'}{(\kf_2+\lf')^2}-\frac{\kf_2}{\kf_2^2} \right)_l   
\hspace{5cm}   
\eeqn   
\beqn   
(T_{il} T'_{jl})^{(c4)}   
=-2    
\left(\frac{\lf - \kf_1 - \kf_2}{D(\lf - \kf_1 - \kf_2)}   
     + \frac{\lf + \kf_1}{D(\lf + \kf_1)}   
     - \frac{\kf_1 + \kf_2}{D(\kf_1 + \kf_2)}   
     +\frac{\kf_1}{D(\kf_1)} \right)_i    
      \left( \frac{\kf_2 -\lf}{(\kf_2-\lf)^2}- \frac{\kf_2}{\kf_2^2}\right)_l   
\nonumber \\   
\cdot \left( \frac{\kf_1+\kf_2}{D(\kf_1 + \kf_2)} -    
                           \frac{\kf_1}{D(\kf_1)} \right)_j   
\left( \frac{\kf_2 +\lf'}{(\kf_2+\lf')^2}-\frac{\kf_2}{\kf_2^2} \right)_l   
\hspace{5cm}   
\eeqn   
\par 
Combining all contributions (a) - (c) we arrive at our final result. The sum    
of the products $T_{il}T'_{jl}$ can be written as the square of two identical   
(up to the primed $l$-integral) expressions:   
\beqn   
T_{il}= \left( \frac{\lf + \kf_1 + \kf_2}{D(\lf + \kf_1 + \kf_2)}   
              + \frac{\kf_1 + \kf_2}{D(\kf_1 + \kf_2)}   
              - \frac{\kf_1- \lf}{D(\kf_1-\lf)}   
              -\frac{\kf_1}{D(\kf_1)}  \right)_i   
        \left( \frac{\lf + \kf_2}{(\lf + \kf_2)^2}   
                  - \frac{\kf_2}{\kf_2^2} \right)_l   
\nonumber \\   
             +\left( \lf \to -\lf \right) \hspace{4cm}.   
\eeqn   
In order to obtain $M_l$, we simply drop the   
numerators in the first factor (with subscript `$i$`) in eq.~(2.38):    
\beqn   
T_{l}= \left( \frac{1}{D(\lf + \kf_1 + \kf_2)}   
              + \frac{1}{D(\kf_1 + \kf_2)}   
              - \frac{1}{D(\kf_1-\lf)}   
              -\frac{1}{D(\kf_1)}  \right)  
        \left( \frac{\lf + \kf_2}{(\lf + \kf_2)^2}   
                  - \frac{\kf_2}{\kf_2^2} \right)_l   
\nonumber \\   
             +\left( \lf \to -\lf \right) \hspace{4cm}   
\eeqn   
which defines $T_l$, and $M_l$ then follows from eq.~(2.24).   
Inserting these results into eq.~(2.12) - eq.~(2.15), we arrive    
at our final expression    
for the $ep$ cross section.   
   
\section{Impact Parameter Representation}\setcounter{equation}{0}  
It is very instructive to analyze the structure of the matrix element $M_{il}$   
or $M_l$ in impact parameter space. As will be shown below the simplification  
in impact parameter space arises in form of two wave functions of which  
the first describes the dissociation of the photon into the $q\bar{q}$ - pair  
and the second the subsequent radiation of the gluon from the quark or   
the antiquark. The couplings of the $t$ - channel  
 gluons generate phase - factors  
which after some regrouping can be turned into effective dipole cross sections.  
For the numerical analysis later on we have, of course, to use  
the momentum representation.  
\par 
We start by taking the Fourier transformation of $T_{il}$:  
\beqn  
\tilde{T}_{il}&=&  
\int d^2\kf_1 \; d^2\kf_2   
\mbox{e}^{i \kf_1 \cdot \rf_1 + i \kf_2 \cdot \rf_2} \;  \;T_{il} \\  
&=&\Psi^T_i(\rf_1)\;\int d^2\kf_2 \; \mbox{e}^{i \kf_2 \cdot \rf_2}  
\left[ \left\{\mbox{e}^{-i (\kf_2 +\lf) \cdot \rf_1}+   
\mbox{e}^{-i \kf_2 \cdot \rf_1}-\mbox{e}^{-i \lf \cdot \rf_1} - 1\right\}  
\left\{ \frac{\kf_2+\lf}{(\kf_2+\lf)^2}-\frac{\kf_2}{\kf_2^2}\right\}_l  
\right. \nonumber \\  
&&\hspace{5cm} +(\lf \rightarrow -\lf)\;]\nonumber\\  
&=& \Psi^T_i(\rf_1)\;\left[ \varphi_l(\rf_2-\rf_1)\;\left\{  
\mbox{e}^{-i \l \cdot \rf_2}+\mbox{e}^{-i \l \cdot (\rf_2-\rf_1)}  
-\mbox{e}^{-i \l \cdot \rf_1}-1\right\} +(\lf \rightarrow -\lf)\right.  
\nonumber\\  
 &&\left.\hspace{1.8cm}-\;\varphi_l(\rf_2)\;\left\{  
\mbox{e}^{-i \l \cdot \rf_2}+\mbox{e}^{-i \l \cdot (\rf_2-\rf_1)}  
-\mbox{e}^{i \l \cdot \rf_1}-1\right\} +(\lf \rightarrow -\lf)\;\right]  
\nonumber  
\eeqn  
where $\Psi^T$ and $\varphi$ are defined as   
\beqn  
\Psi^T_i(\rf)&=&\int d^2\kf \;\mbox{e}^{i \kf \cdot \rf}   
\;\frac{\kf_i}{D(\kf)}\;=\;-2 \pi i\;\frac{\rf_i}{|\rf|}\;  
\sqrt{\alpha(1-\alpha)Q^2}\;\mbox{K}_1(\sqrt{\alpha(1-\alpha)Q^2r^2})  
\nonumber\\  
\varphi_l(\rf)&=&\int d^2\kf \;\mbox{e}^{i \kf \cdot \rf}  
\;\frac{\kf_l}{\kf^2}\;\hspace{0.5cm}=  
\;-2 \pi i\;\frac{\rf_l}{|\rf|^2}\;\;.  
\eeqn  
If we now introduce the effective dipole cross-section as  
\beqn  
\sigma(\rf)&=&\int \frac{d^2l}{\pi l^2} \;{\cal F}(x_{\fP},l^2)\;  
\left(1- \mbox{e}^{i \l \cdot \rf}\right)\;  
\left(1- \mbox{e}^{-i \l \cdot \rf}\right)  
\eeqn  
we finally obtain for $\tilde{M_{il}}$, the Fourier transform of    
$M_{il}$:   
\beqn\label{impact}  
\tilde{M_{il}}&=& \Psi^T_i(\rf_1)\;\left\{ \varphi_l(\rf_2)-  
\varphi_l(\rf_2-\rf_1)\right\}_l\;  
\left[\sigma(\rf_2)+\sigma(\rf_2-\rf_1)-\sigma(\rf_1)\right]\;\;.  
\eeqn  
The vector $\rf_1$ denotes the separation of the quark and antiquark, $\rf_2$  
the separation of the quark and gluon and $\rf_2-\rf_1$   
the separation of the antiquark  
and gluon. The result in eq.~(\ref{impact}) is the same as in   
ref.~\cite{NikZakqqg}. One derives the corresponding  
expression for $\tilde{M}_l$ by substituting the transverse wave function  
$\Psi^T$ by the longitudinal wave function $\Psi^L$,  
\beqn  
\Psi^L(\rf)&=&\int d^2\kf \;\mbox{e}^{i \kf \cdot \rf}   
\;\frac{1}{D(\kf)}\;=\;2 \pi\;  
\mbox{K}_0 \left( \sqrt{\alpha(1-\alpha)Q^2r^2} \right)\;\;.  
\eeqn  
The impact parameter representation suggests the intuitive interpretation that  
following the initial dissociation of the photon,   
which is described by $\Psi^{T,L}$,  
each of the quarks radiates off a gluon represented by $\varphi$.   
After the creation  
of the $q\bar{q}g$-state the interaction with the target processes pairwise,  
i.e. each of the pairs $q\bar{q}$, $qg$ and $\bar{q}g$ gives a separate   
contribution as indicated by the different arguments of the   
effective dipole cross-section $\sigma$.  
As was pointed out in ref.~\cite{NikZakqqg}, however, the color structure  
is not quite consistent with the previous interpretation. In particular the  
interaction of the $q\bar{q}$ as part of the $q\bar{q}g$-final state is   
color   
suppressed by powers of $N_c$.  
The reason for having  
a non-suppressed contribution from the $q\bar{q}$ - pair is hidden  
in the requirement to include   
the interaction of the 'renormalized' $q\bar{q}$ - pair~\cite{NikZakqqg}.  
In other words, the expression in eq.~(\ref{impact})  
also includes a contribution where the interaction with the $q\bar{q}$  
takes place before a gluon is emitted. This contribution is not color   
suppressed. The result eq.~(\ref{impact})   
has been derived from Feynman diagrams  
which automatically take into account all necessary configurations.   
\par 
Taking the square of the amplitude in impact parameter space  
one finds  
\beqn\label{dipole}  
\tilde{M}_{il}\tilde{M}_{il}&=&(2\pi)^2\;|\Psi^T(\rf_1)|^2\;\frac{\rf_1^2}{\rf_2^2\;  
(\rf_2-\rf_1)^2}  
\;\left[\sigma(\rf_2)+\sigma(\rf_2-\rf_1)-\sigma(\rf_1)\right]^2\;\;.  
\eeqn  
The expression $\;\frac{\rf_1^2}{\rf_2^2\;  
(\rf_2-\rf_1)^2}$ is characteristic for the dipole formalism~\cite{MuePat}.   
The exact equivalence of eq.~(\ref{dipole})   
and the corresponding double - dipole scattering still needs to be established.  
\par  
Assuming strong ordering in the separation of the quark - antiquark - pair  
and the separation of the quarks and the gluon, $|\rf_1|\ll |\rf_2|$,  
one obtains the simple factorized form  
\beqn  
\tilde{M}_{il}\tilde{M}_{il}&=&(2\pi)^2\;\rf_1^2\;|\Psi^T(\rf_1)|^2  
\;\;4\;\frac{\sigma^2(\rf_2)}{\rf^4_2}\;\;.  
\eeqn  
  
\section{A Few Special Kinematic Regions}   
\setcounter{equation}{0}   
In this section we consider a few kinematical regions of interest for which    
analytic expressions can be obtained. We will average over azimuthal angles.   
As a result, we need to consider only eq.~(2.12) and eq.~(2.14).  
Eq.~(2.15) would require a discussion similar to eq.(2.14). 
\par   
First we study the `hard` region:    
\beqn   
Q_0^2 \ll \lf^2 \ll \kf_2^2 \sim  D(\kf_1) \sim D(\kf_1+\kf_2)\sim \tilde{\kf}^2   
\eeqn    
where $Q_0^2$ denotes a hadronic scale which separates the hard and the    
soft region. For simplicity we assume that all relevant scales in the problem   
are of the same order of magnitude, $\tilde{k}^2$.   
In the $\gamma^* - \fP$ CM-system the gluon jet is close to the direction    
of the Pomeron, and the two quarks form a broader jet in the opposite    
direction. For $T_{il}$ in eq.~(2.38)   
we find (after averaging over the azimuthal   
angle of $\lf$):   
\beqn   
T_{il}=\frac{\lf^2}{\kf_2^2}   
\left(\frac{\delta_{im}}{D(\kf_1+\kf_2)}-   
               2 \frac{(\kf_1+\kf_2)_i (\kf_1+\kf_2)_m}{D^2(\kf_1+\kf_2)}   
+\frac{\delta_{im}}{D(\kf_1)}-2 \frac{(\kf_1)_i (\kf_1)_m}   
{D^2(\kf_1)}\right)   
\left( \delta_{ml} - 2 \frac{(\kf_2)_m (\kf_2)_l}{\kf_2^2}\right)   
\eeqn   
Using eq.~(2.24) and eq.~(2.25) we obtain for $M_{il}M'_{il}$:   
\beqn   
M_{il}M'_{il}&=&\frac{2}{(\kf_2^2)^2}    
\left\{ \left(\frac{1}{D(\kf_1+\kf_2)} + \frac{1}{D(\kf_1)}\right)^2   
\right. \nonumber \\ &&   
       - 2 \left(\frac{1}{D(\kf_1+\kf_2)} + \frac{1}{D(\kf_1)}\right)   
          \left(\frac{(\kf_1+\kf_2)^2}{D^2(\kf_1+\kf_2)}    
               + \frac{\kf_1^2}{D^2(\kf_1)}\right) \\   
       &&\left. + 2 \left(\frac{(\kf_1+\kf_2)^2}{D^2(\kf_1+\kf_2)}    
               + \frac{\kf_1^2}{D^2(\kf_1)}\right)^2    
       + 4 \frac{[\kf_1\cdot(\kf_1+\kf_2)]^2 - \kf_1^2 (\kf_1+\kf_2)^2}   
                 {D^2(\kf_1+\kf_2) D^2(\kf_1)} \right\}   
       \cdot \left[x_{\fP}g(x_{\fP}, \tilde{\kf}^2)\right]^2 \nonumber   
\eeqn   
The transverse cross section is obtained by inserting eq.~(4.3) into   
eq.~(2.12).   
The most remarkable feature of this result is the appearance of the   
square of the gluon structure function: because of the prefactor    
$1/(\kf_2^2)^2$, one expects the `hard` region of large $\kf_2^2$ to be    
suppressed. But as a result    
of its rise at small $x_{\fP}$, the gluon structure function provides   
an enhancement factor of the hard region which increases with  
decreasing  $x_{\fP}$.   
\par 
As a special case of the hard region we consider the case of strong ordering   
of the transverse momenta, $\kf_2^2 \ll \kf_1^2$.  
Then eq.~(4.3) simplifies into  
\beqn   
T_{il}=2 \frac{\lf^2}{\kf_2^2}   
\left(\frac{\delta_{im}}{D(\kf_1)}-2 \frac{(\kf_1)_i (\kf_1)_m}   
{D^2(\kf_1)}\right)   
\left( \delta_{ml} - 2 \frac{(\kf_2)_m (\kf_2)_l}{\kf_2^2} \right)   
\eeqn   
With the approximation of eq.~(2.19) we obtain for $M_{il}M'_{il}$:   
\beqn   
M_{il}M'_{il}=8\frac{Q^4+m^4}{(Q^2+m^2)^4} \left(\frac{m^2}{\kf_1^2}\right)^2   
\frac{1}{(\kf_2^2)^2}\left[x_{\fP}g(x_{\fP}, \kf_2^2)\right]^2\;\;.    
\eeqn  
It might be interesting to note that an early approach \cite{Ryshera}   
to diffractive deep inelastic scattering which not only considers the radiation of 
one but many gluons was based on the special  
case discussed above. An extension of these results  
towards large $\beta$ can be found in ref.~\cite{LW} which in our notation 
reads 
\beqn 
M_{il}M'_{il}=8\frac{Q^4+m^4}{(Q^2+m^2)^4} \left(\frac{m^2}{\kf_1^2}\right)^2 \; 
\left(\frac{M^2-m^2}{M^2+Q^2}\right)^4\left(\frac{M^2+2 m^2+3 Q^2}{M^2+Q^2}\right)^2\;  
\frac{1}{(\kf_2^2)^2}\left[x_{\fP}g(x_{\fP}, \kf_2^2)\right]^2\;\;.   \nonumber 
\eeqn   
The main observation here is the strong suppression of $M_{il}M'_{il}$  
when $m^2$ becomes of the order of $M^2$.   
This fact is also implemented in the parameterization of \cite{BEKW}.   
\par  
Next we say a few more words about the small-$\kf_2^2$ region.   
As we have noted already after eq.~(4.3),   
the region of small $\kf_2^2$ is expected    
to give an important contribution to the integrated diffractive cross section.   
On the other hand, if $\kf_2^2$ becomes smaller than, say, the hadronic scale   
$Q_0^2$, perturbation becomes unreliable and we cannot apply our formulae.   
For practical purposes, however, it    
would be attractive to find an extrapolation of our   
jet cross section into the region where the gluon jets becomes soft.   
Following the discussion of the $q\bar{q}$   
final state in diffraction dissociation in~\cite{BLW} we   
shall present an extrapolation of our perturbative calculation,     
which at small $\kf_2^2$ leads to   
the Ingelman-Schlein picture of the Pomeron structure function.   
First, at small $\kf_2^2$, the approximation $\lf^2 \ll \kf_2^2$ will no    
longer be valid; $\lf^2$ can become of the same order as or even larger than   
$\kf_2^2$. For simplicity we do not consider the most general case   
but still retain the approximation    
$\lf^2, \kf_2^2 \ll \kf_1^2$.   
After integrating over the azimuthal angle of $\lf$ we find for    
$T_{il}$:   
\beqn   
T_{il}= 2 \left(\frac{\delta_{im}}{D(\kf_1)}-2 \frac{(\kf_1)_i (\kf_1)_m}   
{D^2(\kf_1)}\right)   
\left( \delta_{ml} - 2 \frac{(\kf_2)_m (\kf_2)_l}{\kf_2^2} \right)   
\left(\Theta(\lf^2-\kf_2^2) + \frac{\lf^2}{\kf_2^2} \Theta(\kf_2^2 - \lf^2)   
\right)   
\eeqn   
which is consistent with the result found earlier in ref.~\cite{R}.    
Before inserting eq.~(4.6) into eq.~(2.29),  
 we have to discuss the two gluon amplitude   
${\cal F}(x_{\fP},\lf^2)$. As long as the momentum scale $l^2$ at the upper    
end    
is large, it can (to a good approximation) be identified with the unintegrated   
gluon structure function (see eq.~(2.25)).  
 Now we will try to find an extrapolation   
into the soft $\kf_2^2$ region which, at low $\kf_2^2$, smoothly turns into   
the Pomeron structure function picture. An easy way has been    
sketched in~\cite{BLW}, and here we outline the argument in somewhat    
more detail.   
We make the following simple double Mellin transform ansatz for ${\cal F}$:   
\beqn   
{\cal F}(x_{\fP},\lf^2) = \phi_0 \frac{1}{Q_0^2}    
\int \frac{d\omega}{2\pi i} \int \frac{d\mu}{2\pi i}   
\left(\frac{1}{x_{\fP}} \right)^{\omega}   
\left( \frac{\lf^2}{Q_0^2} \right)^{\mu}   
\frac{1}{\omega - \chi(\mu)}   
\eeqn   
where the integration contours run along the imaginary axis    
($-1 < Re \mu <0$, $\chi(\mu) < Re \omega$), and $\phi_0$ denotes the overall   
normalization constant.   
If we would perform the $\mu$ integral first by picking up a pole at   
$\mu=\chi^{-1}(\omega)=\gamma(\omega)$, we obtain the anomalous dimension.   
However, here it is more convenient to   
model the function $\chi(\mu)$ and to do first the $\omega$ integral.   
Let us assume that $\chi(\mu)$ is positive and becomes large both at   
$\mu=-1$ and $\mu=0$ (similar to the BFKL characteristic function).   
We then obtain:   
\beqn   
{\cal F}(x_{\fP},\lf^2) = \phi_0  \frac{1}{Q_0^2}    
\int \frac{d\mu}{2\pi i}   
\exp \left (\chi(\mu) \log \frac{1}{x_{\fP}} +   
\mu \log \frac{\lf^2}{Q_0^2} \right)   
\eeqn   
Combining this ansatz with eq.~(4.6) we write the $l^2$ integral as:   
\beqn   
\left( \frac{\kf_2^2}{Q_0^2} \right)^{\mu}  \int^{\kf_1^2/\kf_2^2}    
\frac{d\lf'^2}{\lf'^2} \left( \lf'^2 \right)^{\mu} \left[\Theta(\lf'^2 -1)   
+\lf'^2 \Theta(1-\lf'^2)\right]   
\eeqn   
Inserting the $\kf_2^2$-dependent prefactor of eq.~(4.9)   
into the $\mu$ integral of   
eq.~(4.7) and performing a stationary phase analysis, we find that   
for $Q_0^2 \ll \kf_2^2$ the stationary point moves towards $\mu=-1$, whereas   
for the opposite case $\kf_2^2 \ll Q_0^2$ the stationary point is near $\mu=0$.   
In order to evaluate the $\lf'^2$ integral in eq.~(4.8)  
 near $\mu=-1$ or $\mu=0$   
we have to specify the lower limit of integration.   
In the first case (large $\kf_2^2$, $\mu \to -1$), the first term in eq.~(4.9)    
stays constant    
whereas the second part gives a $\log \kf_2^2 / Q_0^2$ from the region    
$Q_0^2/\kf_2^2 < \lf'^2 < 1$. In the second case (small $\kf_2^2$, $\mu \to 0$)    
the first term gives a logarithmic enhancement $\log \kf_1^2/Q_0^2$ whereas    
the second   
term stays constant (which is equal to 1 if the $l'^2$ integral extends    
down to 0). As a result we obtain for the $\lf^2$ integral:   
\beqn   
\Phi(\kf_1^2, \kf_2^2,x_{\fP})&=& \int^{\kf_1^2} \frac{d\lf^2}{\lf^2}    
{\cal F}(x_{\fP},\lf^2)   
\left(\Theta(\lf^2-\kf_2^2) + \frac{\lf^2}{\kf_2^2} \Theta(\kf_2^2 - \lf^2)   
\right) \nonumber \\   
&\sim& \left\{  \begin{array}{ccc}   
\phi_0/\kf_2^2\;\ln\left( \kf_2^2/Q_0^2\right) &  
  \kf_2^2 \gg Q_0^2 &    
\mu \sim -1 \\ \phi_0/Q_0^2\;\ln \left(\kf_1^2/\kf_2^2\right) &   
 \kf_2^2 \ll Q_0^2 & \mu \sim 0   
     \end{array} \right\}   
\eeqn   
Together with eq.~(4.6), we obtain for $M_{il}M'_{il}$:   
\beqn   
M_{il}M'_{il}= \Phi^2 \frac{8m^4}{(\kf_1^2)^2} \frac{Q^4 +m^4}{(Q^2+m^2)^4}   
\eeqn   
Inserting this into (2.13), and making use of (2.20) we find the expected   
result: the transverse cross section belongs to leading twist, and the   
integral over $\kf_1^2$ diverges logarithmically. 
\par  
A nice feature of this extrapolation into the soft non-perturbative region   
is the following. As we have described above, the two cases (large and small   
$\kf_2^2$) correspond to different regions of the $\lf^2$ integral: the first case   
is the usual strong ordering with $Q_0^2 < \lf^2 < \kf_2^2$. In the second   
case $\lf^2$ prefers to be larger than $\kf_2^2$: in this region we simply   
drop $\kf_2^2$ in comparison with $\lf$ and $\kf_1$. As a result, the non-planar   
diagrams in Fig.~\ref{diag} become small in comparison with the planar ones.   
What is left can be interpreted as a `gluon component of the Pomeron    
structure function`. This has to be compared with the first case, where   
all diagrams are equally important, i.e. the Pomeron interacts with the    
whole diffractive system. 
\par  
Next we consider the case where the gluon   
stays hard (large $\kf_2$) and one of the quarks    
becomes soft, i.e. $\kf_1\sim 0$ or $\kf_1+\kf_2\sim 0$.   
Let us assume $\kf_1$ to be small ($\kf_1^2<\lf^2<Q_0^2$)    
and go back to eq.~(2.38).   
With $\lf^2>\kf_1^2$ only the third term in the first bracket contributes and we   
find after integration over the azimuth angle:   
\beqn    
T_{il}=    
\left( \delta_{il} - 2 \frac{(\kf_2)_i (\kf_2)_l}{\kf_2^2} \right)   
\frac{\lf^2}{D(\lf)}\;\;.   
\eeqn   
Using our ansatz for the unintegrated structure function eq.~(4.7) with    
the soft extrapolation $\mu \rightarrow 0$ we can perform the     
$\lf$-integration   
\beqn   
\Phi(k_1,x_{\fP})&=& \int^{Q_0^2}_{\kf_1^2} \frac{d\lf^2}{\lf^2}    
\left(\frac{\lf^2}{Q_0^2}\right)^{\mu}\frac{\lf^2}{D(\lf)}   
\nonumber \\   
&\sim& \ln \left(\frac{D(Q_0)}{D(\kf_1)}\right)   
\eeqn   
and finally arrive at   
\beqn   
M_{il}M'_{il}=\Phi^2 \frac{2}{(\kf_2^2)^2}\;\; .   
\eeqn   
It is important to note that this kinematical limit yields    
a contribution which breaks (collinear) factorization. The result in   
eq.~(4.14) is similar to    
the 'super-hard component of the Pomeron' introduced in   
ref.~\cite{FrankStrik}.   
In a scenario where the photon virtuality $Q^2$ is    
smaller than the gluon momentum $\kf_2^2$, one would expect from the quark loop   
a large logarithm of the form $\log(\kf_2^2/Q^2)$, if    
factorization worked. The potential softness of one of the quarks   
which controls the scale $\lf^2$ in eq.~(4.13) is responsible for the lack   
of such a hard logarithm. In other words the soft part of the process    
decouples from the jets in the final state. With regard to the inclusive   
diffractive cross section, the hard gluon jet combined with a soft   
quark gives only a rather small fraction of the total contribution.    
The leading contribution comes from the scenario discussed earlier    
where the gluon becomes soft and the quarks are hard. In this case   
we have the familiar strong ordering situation which provides a hard   
logarithm $\log(Q^2/\kf_2^2)$ in accordance with factorization for diffractive   
deep inelastic scattering~\cite{C} and the notion of a    
Pomeron structure function. Factorization is still    
violated by contributions with $\kf_2^2$ bigger than the assumed    
factorization scale. Since the mean value of $\kf_2^2$ is close to    
$Q_0^2$, as can be inferred from eq.~(4.10) and eq.~(4.11), one can    
argue that for any scale well above $Q_0^2$, factorization should   
hold.  
\par   
Finally we return to our `hardness condition` in eq.~(2.2).   
For the gluon jet, its transverse momentum square $\kf_2^2$ coincides   
with the virtuality of the vertical gluon line in Fig.~\ref{diag}$a$: for small  
$\kf_2^2$ this `Pomeron structure function` diagram becomes more important     
than the other non-planar diagrams; the boundary between hard and soft physics    
is defined by the off-shellness of the gluon emitted from the $q\bar{q}$ pair.    
Correspondingly, for the quark-antiquark pair it is the virtuality of the   
quark (or antiquark) before it emits the gluon. With the approximations   
discussed in the beginning of this section, the requirement that the   
four momentum square of this quark has to be large then leads to the   
last two equations in eq.~(2.2).

\section{A Numerical Study}   
In this section we study the specific signature of the process  
$e + p \to q + q + g + p'$   
with the cross section calculated in the previous sections. We work   
in the $\gamma$ - $\fP$ center of mass system, and  
the gluon will be restricted to the forward hemisphere of the   
proton (Pomeron). We have implemented   
the formulae of eq.~(2.12) - eq.~(2.17) into a full hadron level Monte Carlo    
program~\cite{RAPGAP}. To avoid   
soft divergences of the matrix element we impose a cut in    
$|\hat{t}|=|(q-p_{q} )^2| > 2 $ GeV$^2$ and    
$|\hat{u}|=|(q-p_{\bar{q}} )^2| > 2 $ GeV$^2$,    
where $q$, $p_{q}=q+r-k_1-k_2$, and $p_{\bar{q}}=k_1$ are the four momenta of the    
photon, the upper outgoing quark and the lower outgoing antiquark  
(Fig.~\ref{Kin}),    
respectively, and a cut on the transverse momentum squared of the   
gluon $\kf_2^2 > 2$ GeV$^2$. The coupling $\alpha_s$ is kept fixed at   
$\alpha_s=0.25$. For the gluon structure function we use   
the NLO parameterization of GRV~\cite{GRV}.   
We divide this numerical study into two parts. In order to analyze the    
general features of the three-parton final state we first work in an    
``ideal`` (i.e. truly asymptotic) kinematic environment where  
the invariant mass $M$  is much bigger than all transverse momenta  
($Q^2=100$ GeV$^2$,    
$M^2=900$ GeV$^2$, (i.e.$\beta=0.1$), $x_{\fP}<10^{-2}$).    
In the second part we turn to HERA kinematics.   
\subsection{Asymptotic region}   
First we have to define the region of validity of our calculation.   
In deriving our cross section formulae we have made several approximations.   
Regge-kinematics requires that $\alpha_2 \ll \alpha_1$, $1-\alpha_1$.   
In our numerical analysis we therefore impose a kinematical cut and require   
that $\alpha_2 < \frac{1}{5}\alpha_1$, $\frac{1}{5}(1-\alpha_1)$.   
Second, we expect that for large $M^2$ the mass of the $q\bar{q}$ system    
will be small: in order to enforce this condition we further demand that   
$m_{q\bar{q}}^2 < \frac{1}{4} M^2$.\\  
\begin{figure}[htb]  
\begin{center}  
\epsfig{figure=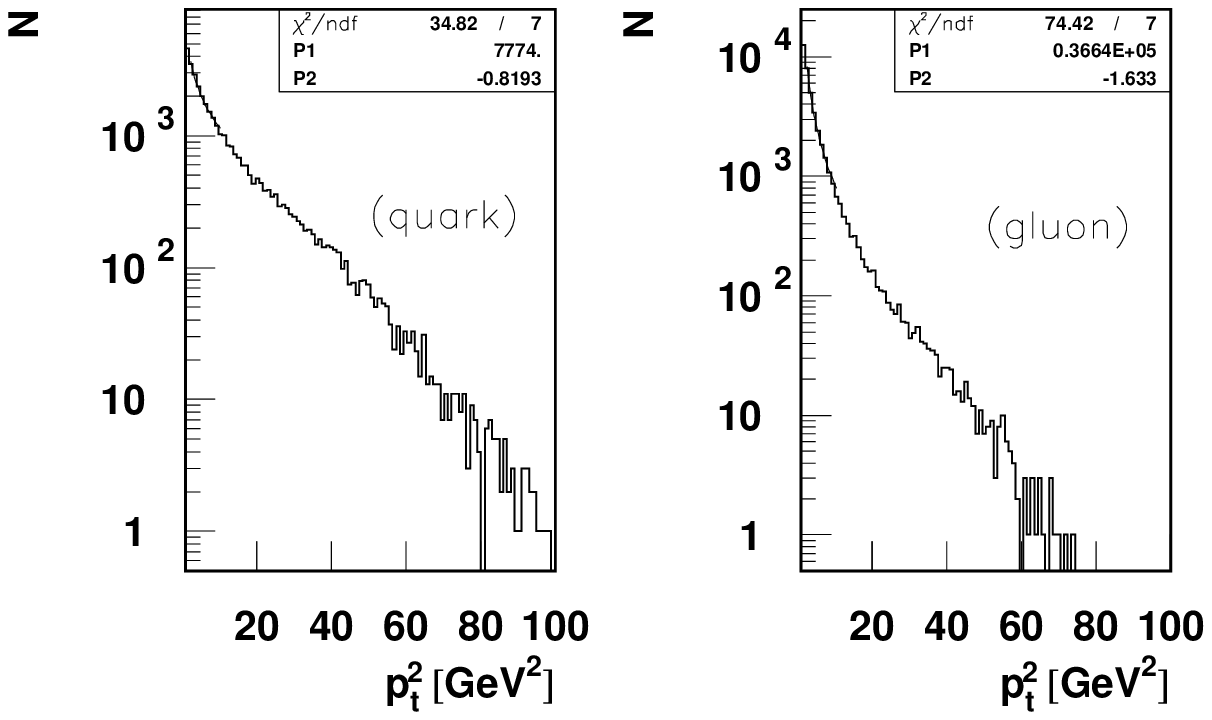}   
\end{center}   
\captive{The distribution of transverse momenta for    
quarks and gluons. \label{pt_asym}}  
\end{figure}\\  
We begin with $p_T$ spectra (Fig.~\ref{pt_asym}). A fit to a power-like behavior   
$(1/p_T^2)^p$ (for $p_T^2< 10$ GeV$^2$)    
gives $p=0.82$ and $p=1.63$ for the quark and gluon transverse   
momenta, respectively.   
This behavior is not far from the naive estimate based upon    
eq.~(4.5): together with the phase space factors in eq.~(2.12) - eq.~(2.13) one expects   
$p=1$ for the quark momentum $\kf_1^2$, and $p$ slightly less than $2$ for the   
gluon momentum $\kf_2^2$. A very striking feature is the fact that for a    
rather large fraction of   
events the ordering condition $\kf_2^2 < \kf_1^2$, $(\kf_1+\kf_2)^2$ is not    
fulfilled:   
we find that only $60 \%$ of the events satisfy this condition.   
This demonstrates that the simple boson-gluon picture in which the   
gluon is emitted with a transverse momentum much smaller than that of the two   
quarks may be very unreliable. On the other hand, for almost all events   
the square of the gluon transverse momentum is less than the virtualities of    
the exchanged quarks ($\kf_2^2 < \hat{|t|}$, $\hat{|u|}$).   
\begin{figure}[htb]  
\begin{center}  
\epsfig{figure=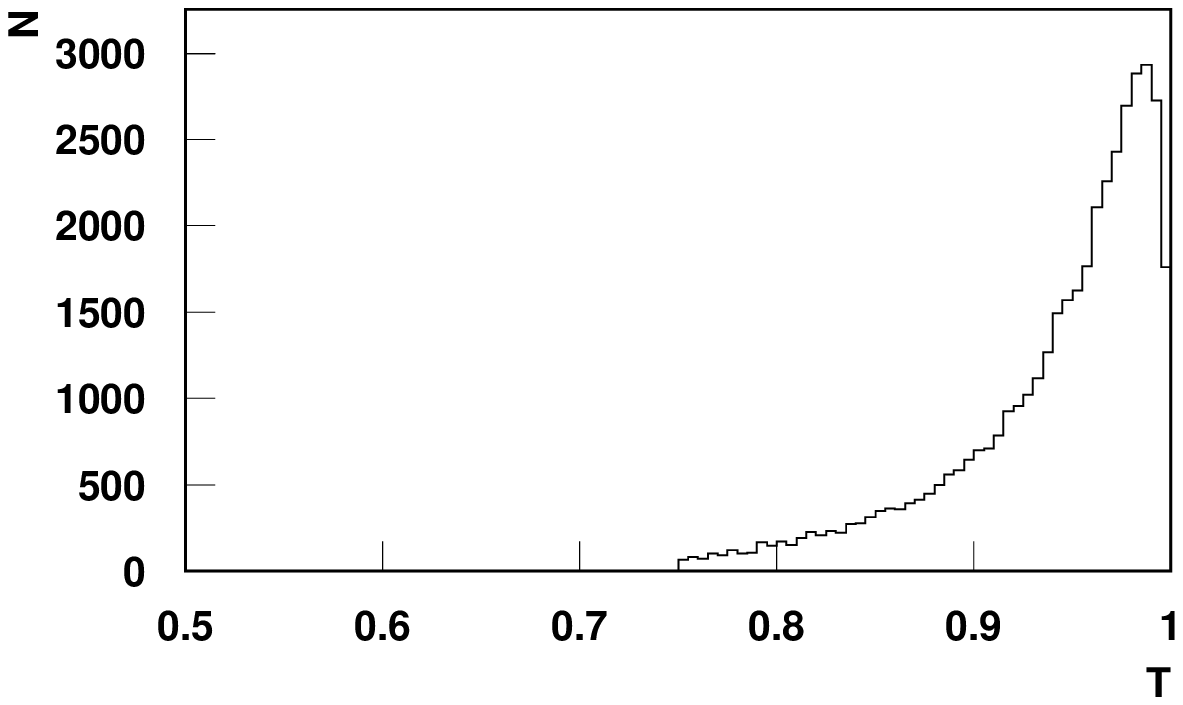}  
\end{center}  
\captive{The distribution of thrust $T$.  
\label{thrust_asym}}  
\end{figure} 
\par  
Next we look into the spatial distribution of the three-parton final state.   
From the experimental point of view it is convenient to begin with the thrust   
distribution: the plot in    
Fig.~\ref{thrust_asym}    
shows that a large fraction of the events    
has a two-jet like structure ($T>0.9$). However, there remains also a sizable    
fraction (about $20\%$) of three-jet events for which $T<0.9$.   
\begin{figure}[htb]   
\begin{center}   
\epsfig{figure=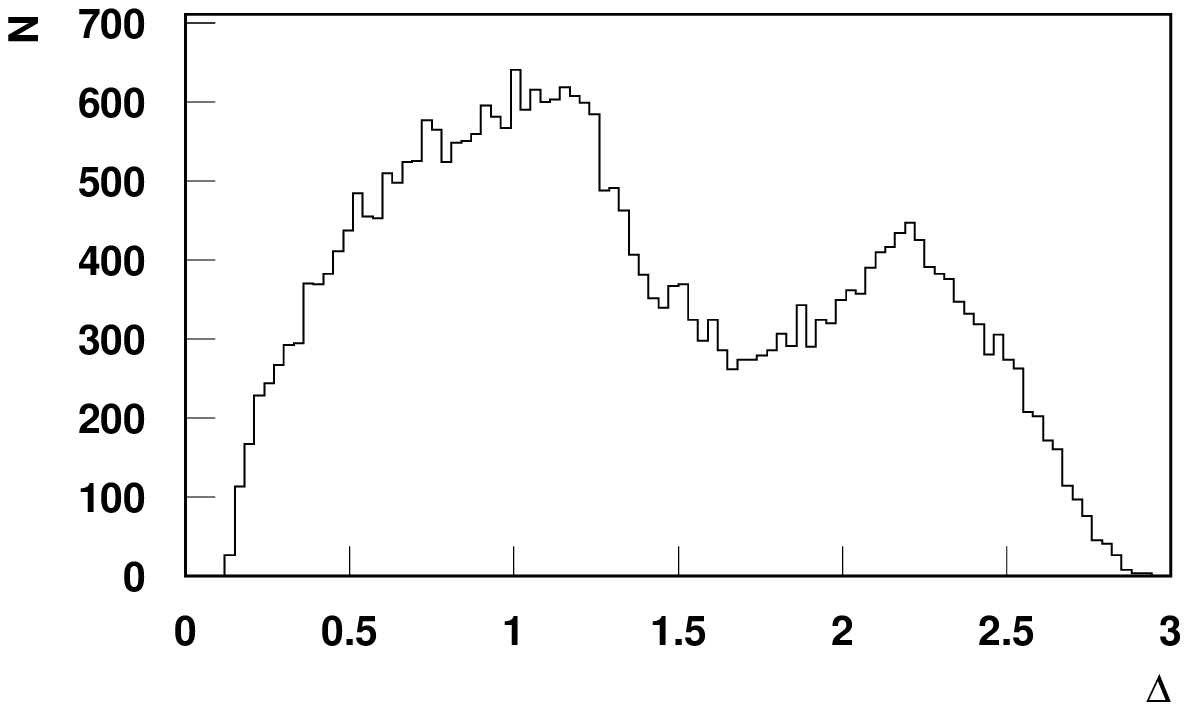}   
\end{center}   
\captive{The distribution of the opening angle $\Delta$ of the $q\bar{q}$ pair  
for values of thrust $T>0.9$ (two jet configuration).  
\label{opangle_asym}}  
\end{figure}  
 To analyze these samples   
in more detail, we begin with the two-jet events and we show in    
Fig.~\ref{opangle_asym} the    
distribution of the opening angle $\Delta$ of the $q\bar{q}$ pair: the    
distribution peaks at both small and large opening angles.   
\begin{figure}[htb]  
\begin{center}  
\epsfig{figure=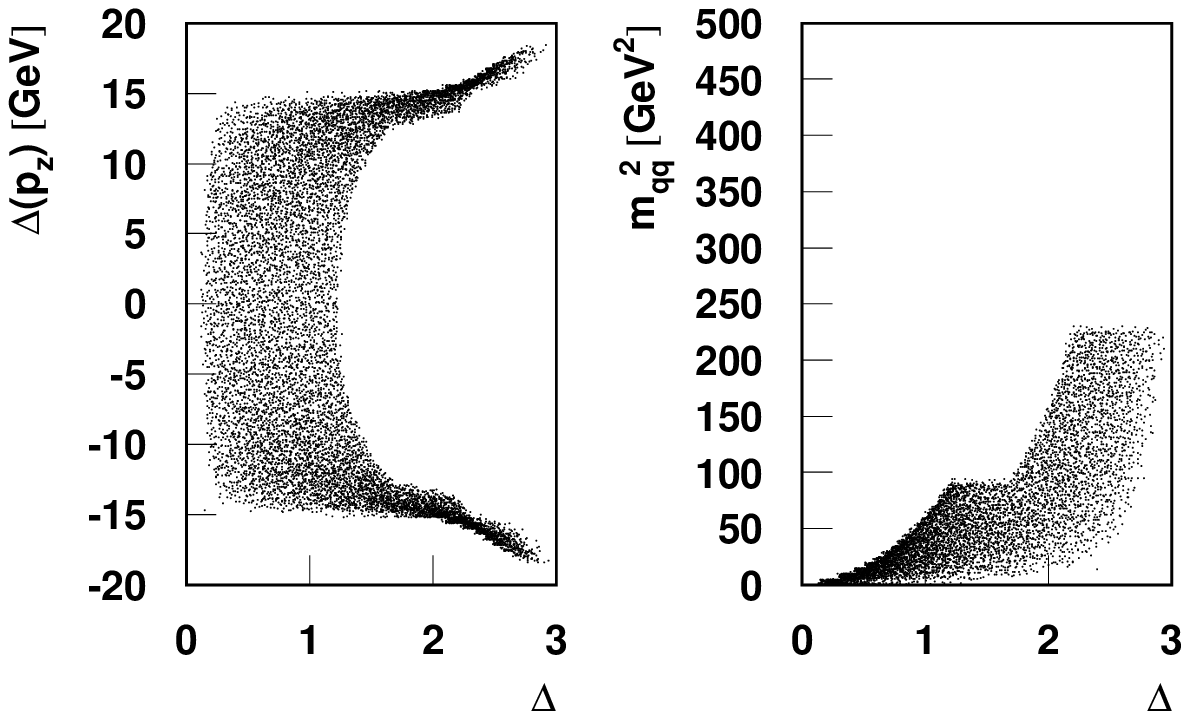}   
\end{center}   
\captive{The difference of the longitudinal   
momenta of the two quarks along the z direction  
($\Delta(p_z)$) as a function of the opening angle $\Delta$ of the $q \bar{q}$  
pair for thrust $T>0.9$. The second plot shows the correlation of the   
$q\bar{q}$-invariant mass and $\Delta$.  
\label{pz_asym}}  
\end{figure}  
 We divide the events into two sets: events with   
large opening angle ($\Delta >\pi/2$) and those with small opening angle   
($\Delta <\pi/2 $). In the first case we find    
(Fig.~\ref{pz_asym}) a large asymmetry in the longitudinal momenta and a big  
invariant mass of the two quarks.  
 \begin{figure}[htb]  
\begin{center}  
\epsfig{figure=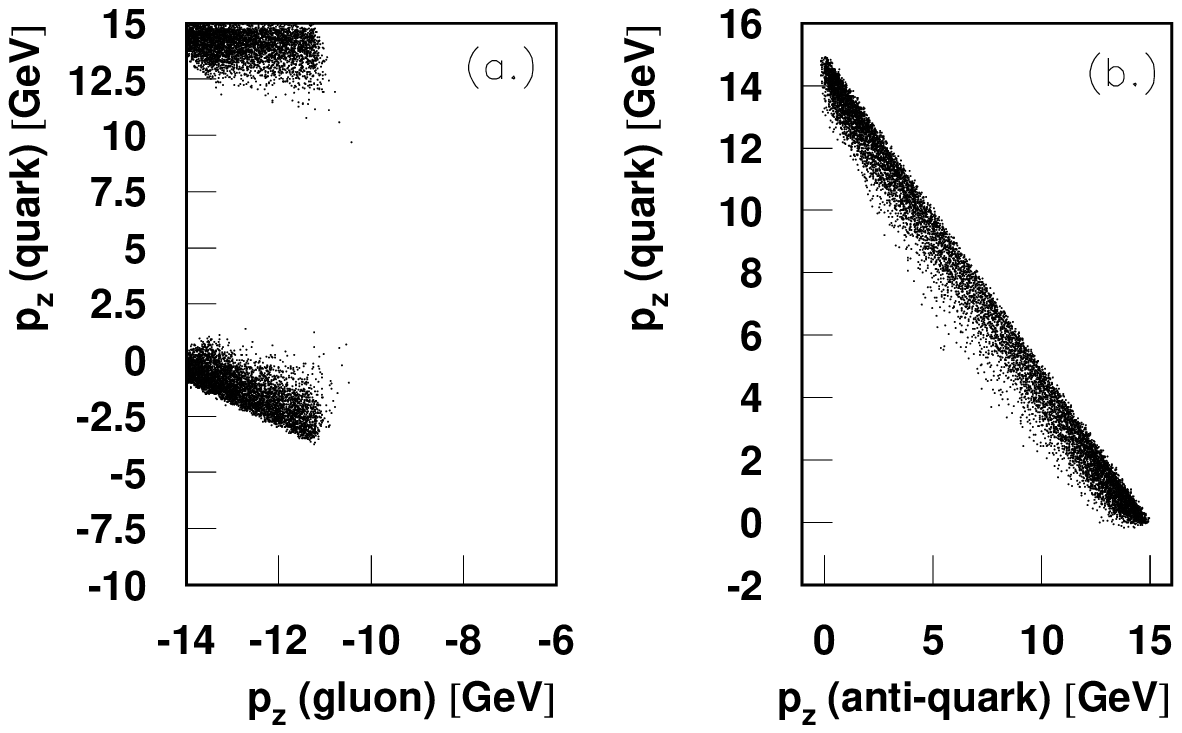}   
\end{center}   
\captive{$a.$ The correlation of the longitudinal  
quark and gluon momentum for $T>0.9$ and $\Delta>\pi/2$ (large opening angle).   
$b$. The longitudinal momenta of  
 the quark and antiquark for $T>0.9$ and $\Delta<\pi/2$ (small opening angle).  
  Both momenta are   
strongly correlated.  
\label{pzpz_asym}}  
\end{figure}  
The quark which moves with maximum momentum   
($\sim$ 15~GeV) in the photon direction forms one jet whereas the second  
quark moves into the opposite direction. The second quark and the gluon   
form the second jet sharing the jet - momentum.  
In Fig.~\ref{pzpz_asym}$a$  
the correlation of the longitudinal  
quark and gluon momentum for $T>0.9$ and $\Delta>\pi/2$ (large opening angle)  
is shown.   
The upper band contains events where the quark moves isolated in the photon direction  
whereas the antiquark moves along with the gluon in the Pomeron direction.   
The lower band contains    
events where the quark and the gluon form a jet sharing the jet-momentum.  
\begin{figure}[htb]   
\begin{center}   
\epsfig{figure=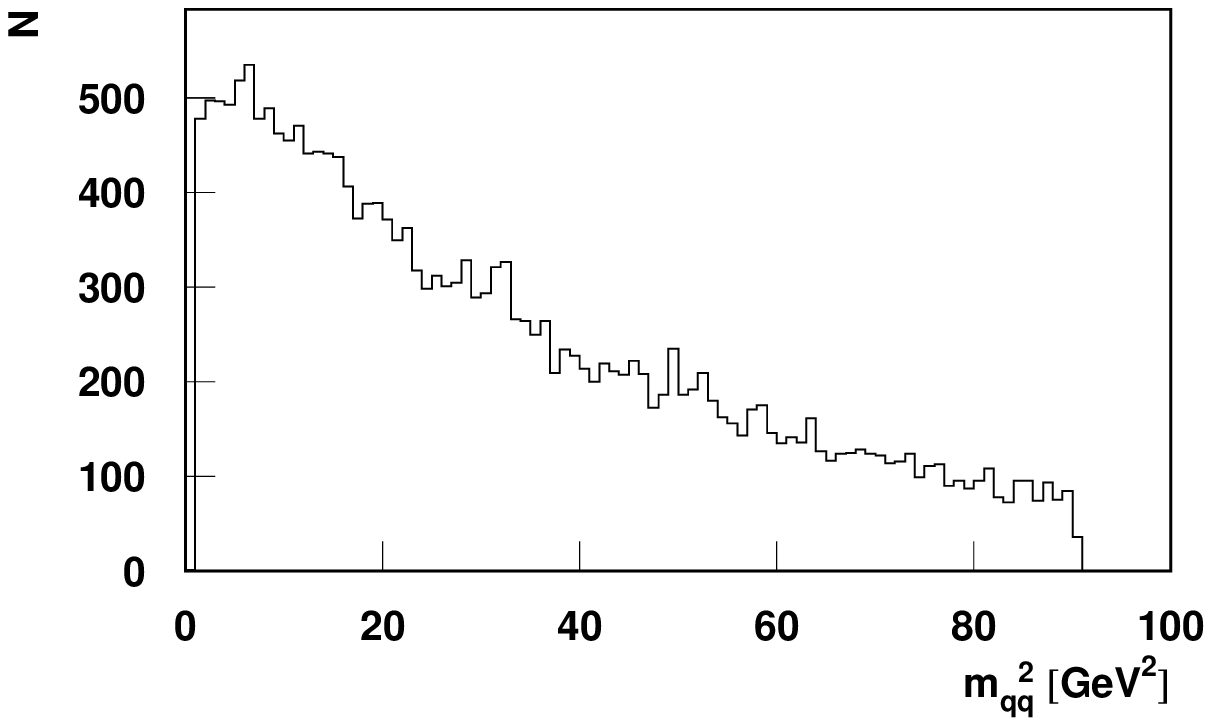}   
\end{center}   
\captive{The distribution of the mass $m_{q\bar{q}}^2$ of the $q \bar{q}$ pair   
 for thrust $T>0.9$  
\label{m2_asym}}  
\end{figure}   
\par  
For those events with small opening angle, on the other hand, the momenta of    
the two quarks are much more symmetric. In this case, the two quarks form a   
jet opposite to the gluon (Fig.~\ref{pz_asym}). 
\par 
In Fig.~\ref{pzpz_asym}$b$ we 
show the quark and antiquark longitudinal   
momentum for $T>0.9$ and $\Delta<\pi/2$ (small opening angle). Both  
momenta are strongly correlated.  
\begin{figure}[htb]   
\begin{center}   
\epsfig{figure=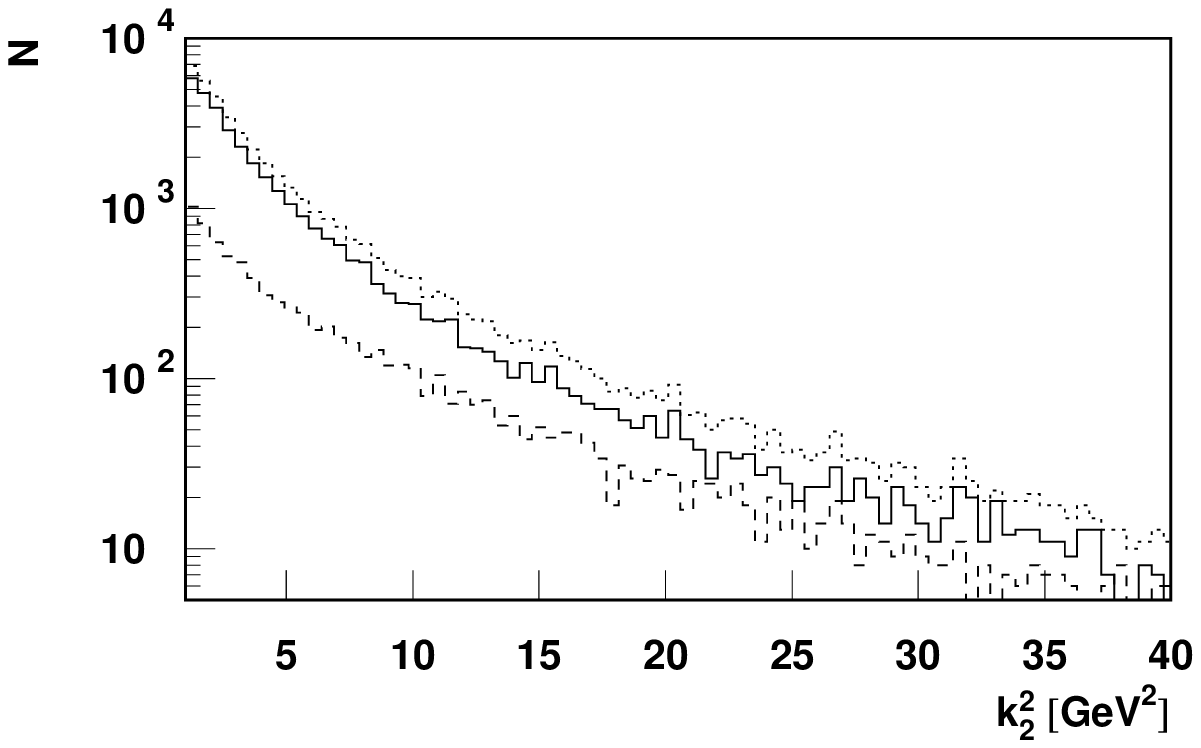}   
\end{center}   
\captive{The transverse momenta  of the gluon for events with $T>0.9$    
(solid line), for $T<0.9$ (dashed line) and for all events (dotted line).   
\label{pt_23jet}}   
\end{figure}   
\par  
In all these cases the jet axis (i.e. the direction of the gluon or quark)   
lies mainly in the Pomeron direction (''aligned gluon configuration''): 
 Fig.~\ref{pt_23jet}  (full curve) shows  
the $\kf_2^2$ distributions of the two - jet events.  
Compared to the full set of events (Fig.~\ref{pt_23jet} dotted curve)  
one notices a  slightly steeper decrease.  
\begin{figure}[htb]   
\begin{center}   
\epsfig{figure=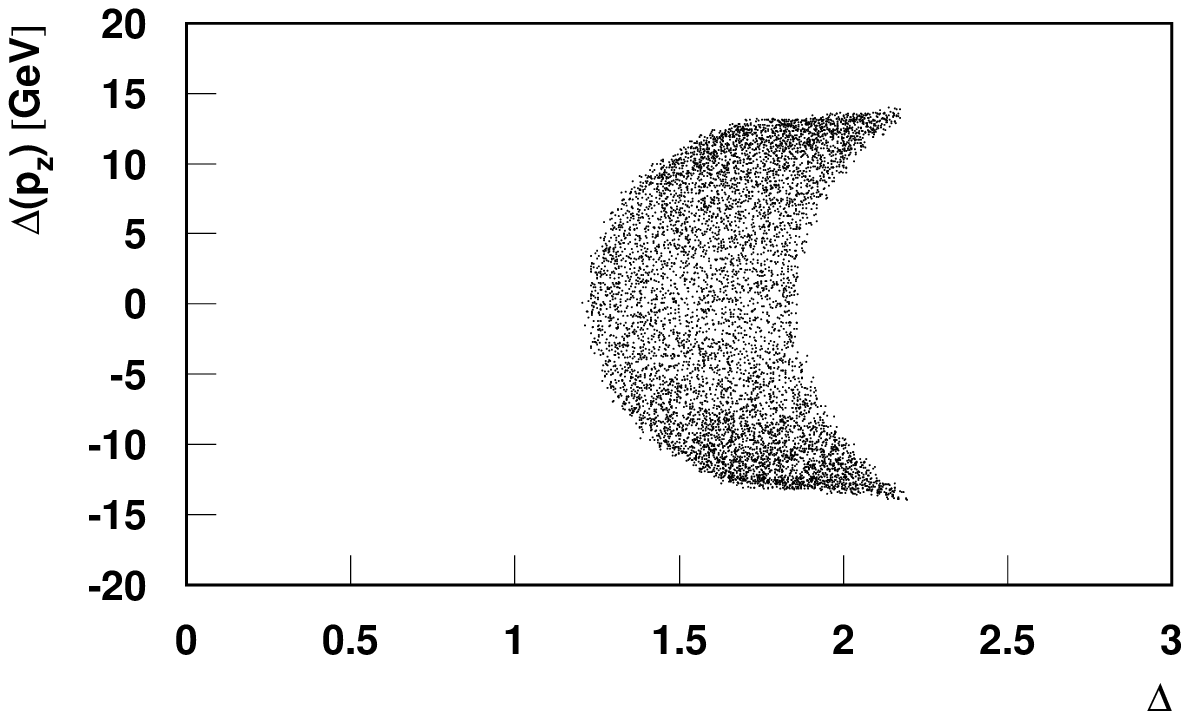}   
\end{center}   
\captive{The difference of the momenta of the two quarks along the z direction   
($\Delta(p_z)$) as a function of the opening angle $\Delta$ of the $q \bar{q}$   
pair for thrust $T<0.9$.   
\label{pz_asym_Tgt09}}   
\end{figure}  
\par  
For the three jet events ($T<0.9$) the opening angle $\Delta$   
ranges between $1.2$ and $2.2$   
(Fig.~\ref{pz_asym_Tgt09}), i.e. we have a star - like configuration.   
Again, the gluon prefers to be in the forward direction    
(Fig.~\ref{pt_23jet}, dashed curve).   
However, compared to the two - jet events, the fall - off at large $\kf_2^2$ is    
less steep. In other words, events with a large gluon transverse momentum   
are more likely to be three - jet like than those with a small transverse    
momentum.   
A few numbers: for $\kf_2^2< 10$ GeV$^2$ the ratio of 3 - jet to 2 - jet  
 events is   
about $0.2$, whereas for $\kf_2^2 > 10$ GeV$^2$ this ration changes to $0.47$.   
\subsection{HERA Region}   
After this more general study we now turn to the HERA region. The results of    
the previous study should apply as long as the diffractive mass is   
sufficiently large:   
only under these conditions we have a sufficiently large   
range in transverse momenta. In HERA kinematics this requirement   
forces us into the (difficult) region of very small $\beta$:    
for $Q^2=5$ GeV$^2$ and $M^2=500$ GeV$^2$, we would need $\beta=0.01$.   
In a first step we have repeated our above analysis for this region, and   
we have found that, indeed, the previous conclusions on our model still apply.   
In this region, however, we expect secondary exchanges to become important,   
and the theoretical interpretation of a jet analysis becomes more    
complicated. 
\par  
A more realistic $\beta$-region is $\beta=0.1$. In the HERA region of not too   
large $Q^2$ we are then limited in $M^2$. This restriction, together   
with the lower cutoffs on the transverse momenta and the virtualities of the   
partons, lead to a severe limitation of the phase space for the    
transverse momenta of the jets. In our numerical analysis we have chosen   
$Q^2=5$ GeV$^2$ and $M^2=45$ GeV$^2$.  We do not impose a cut on $m_{q\bar{q}}^2$ or  
$\alpha_1$ and $\alpha_2$. Such a cut would suppress the cross section by   
a factor of around 10 rendering the analysis meaningless. Our strategy  
will be a comparison with the asymptotic  
situation. From this we can deduce how far we can trust our   
asymptotic formulae for the HERA kinematics. 
\par   
\begin{figure}[htb]   
\begin{center}   
\epsfig{figure=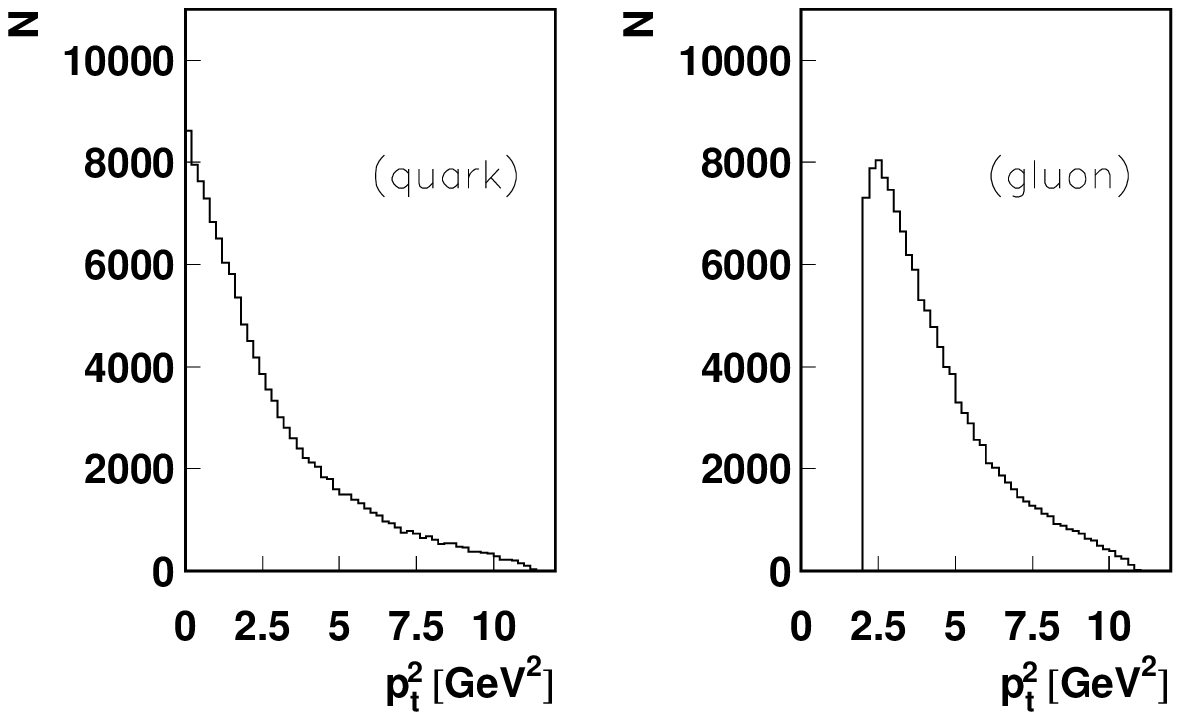}   
\end{center}   
\captive{The distribution of transverse momenta for    
quarks and gluons.   
\label{pt_hera}}   
\end{figure}
\begin{figure}[htb]   
\begin{center}   
\epsfig{figure=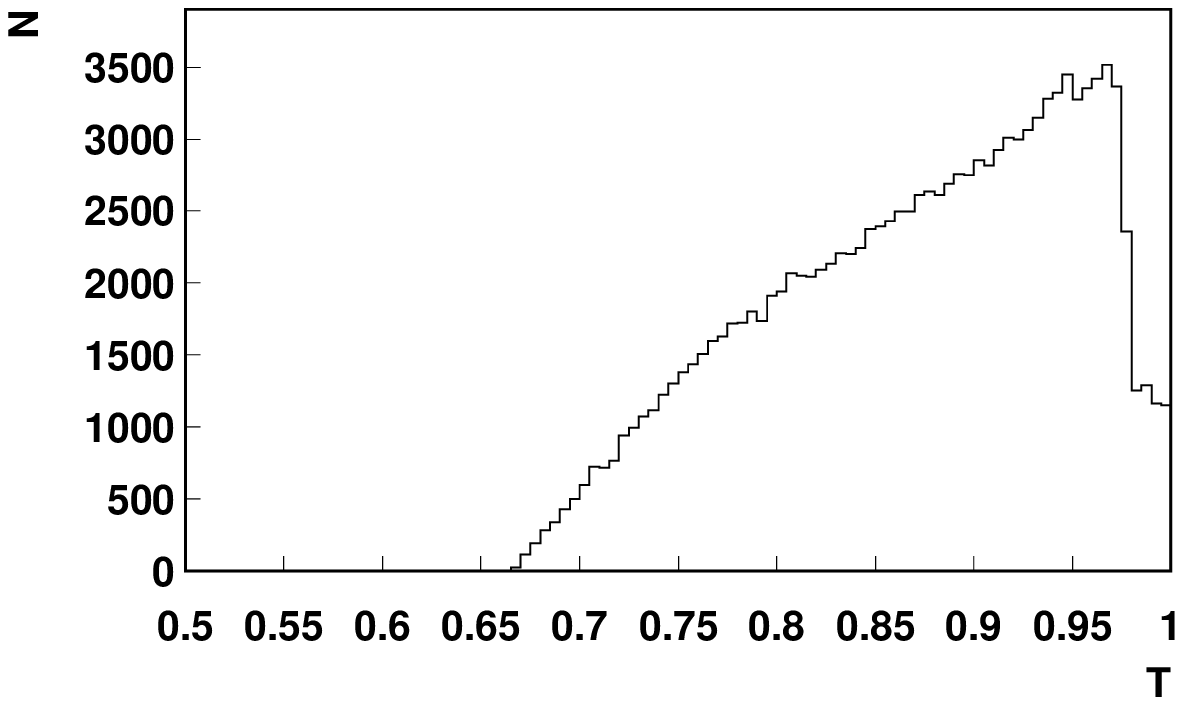}  
\end{center}   
\captive{The distribution of thrust $T$ at HERA energies.   
\label{thrust_hera}}  
\end{figure}

  
We again begin with the $k_t$-spectra (Fig.~\ref{pt_hera})  
for the proposed HERA kinematics,  
and we fit the power $p$. For the quark-spectrum we find  
$p=1.8$ and for the gluon-spectrum $p=2.3$. Both distributions have become  
steeper as compared to the asymptotic case which is mainly due to the   
stronger restriction in the phase space.   
The available energy is approximately a factor  
10 smaller now. We also learn that a measurement of the $k_t$-spectrum  
cannot directly be related to the analytic estimates we have done in the   
previous section.  
In order to test to what extent the usual assumption of strong   
ordering between the transverse momenta of quarks and gluons is satisfied   
we compute the fraction of the cross section where $\kf_2^2 <$    
min$(|\hat{t}|, |\hat{u}|)$: we find that approximately only $2/3$ of the   
cross section satisfies this constraint, whereas $1/3$ lies outside this    
region.   
Next we have a look at the thrust distribution (Fig.~\ref{thrust_hera}).  
It is much broader than in the asymptotic case. Since $M^2$ is no longer  
that large, the relative proportion of transverse to  
longitudinal momenta is much higher than before, and   
therefore leads to an increase of three-jet events (around 60\% of the events  
have a thrust smaller than 0.9).   
\begin{figure}[htb]   
\begin{center}   
\epsfig{figure=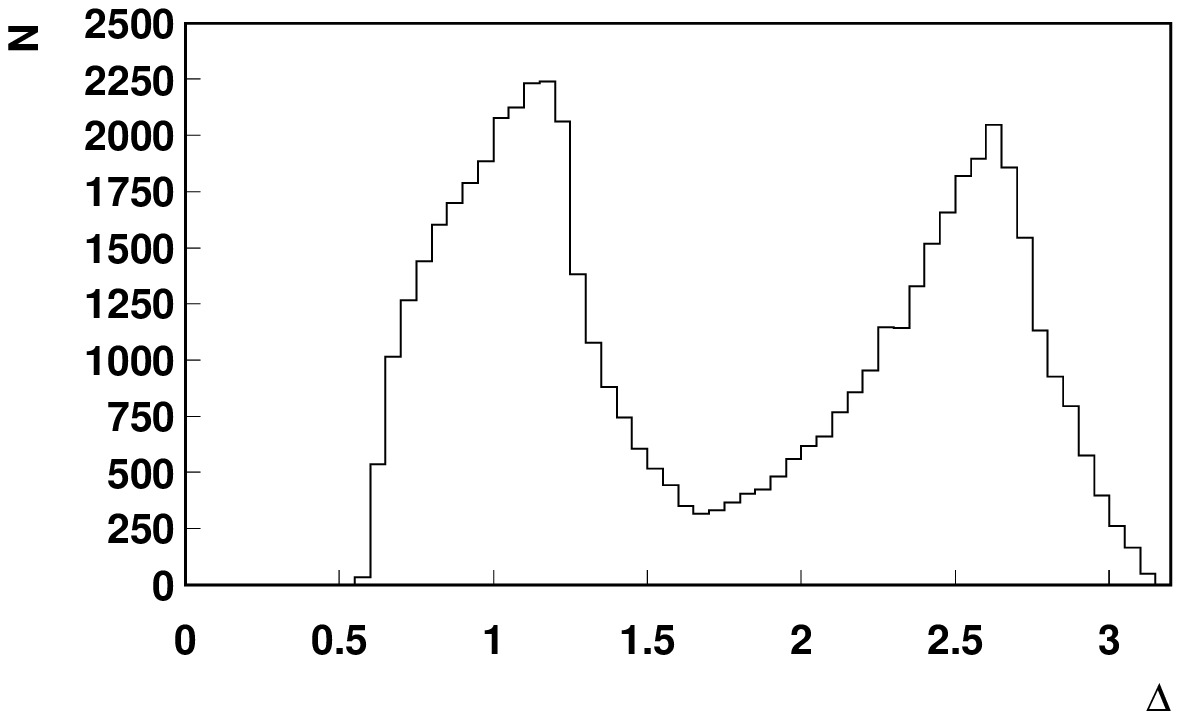}   
\end{center}   
\captive{The opening angle $\Delta$ for $T>0.9$ at HERA energies.  
\label{delta_hera}}  
\end{figure}   
If we ask for two-jet events, i.e. thrust $T>0.9$, then we find again   
two configurations (Fig.~\ref{delta_hera}), one with rather small   
opening angle between the quarks and a second  
with large opening angle. The distribution of $\Delta p_z$ versus opening  
angle $\Delta$ looks similar to the corresponding plot for asymptotic   
energies which makes us believe that the asymptotic formulae also work properly  
for HERA kinematics. 
\par   
\begin{table}   
\begin{center}   
\begin{tabular}{|c|c|c|} \hline   
 kinematic cut & dijet & $q\bar{q}g$ \\ \hline \hline   
     2 GeV$^2$ & 117 pb & 66 pb\\   
     5 GeV$^2$ & 55 pb & 9.2 pb \\ \hline   
\end{tabular}   
\captive{Integrated cross sections for diffractive $q\bar{q}$ and $q\bar{q}g$   
production in the kinematic region defined by:  
$10$~GeV$^2$ $< Q^2$, $50 < W < 220$~GeV, $x_{\fP} < 10^{-2}$. The $t$  
dependence is assumed to be exponential and the $t$ integral is performed. The  
kinematic cut requires all parton $k_t^2$ and   
in the case of $q\bar{q}g$ also $|\hat{t}|,|\hat{u}|$ to be  
greater than the value specified.  
}   
\end{center}   
\end{table}  
Table 1 shows the integrated    
$q\bar{q}g$ diffractive jet cross sections.   
For comparison, we also present dijet cross sections from $q\bar{q}$   
production. We have integrated over the kinematic region   
$10$~GeV$^2$ $< Q^2$, $50 < W < 220$~GeV, $x_{\fP} < 10^{-2}$, where we   
have integrated over $t$, assuming an exponential $t$ dependence.   
In the first row we present cross sections for the case where   
all parton $k_t^2$ are larger than $2$~GeV$^2$ in addition to $|\hat{t}|,   
|\hat{u}| > 2$~GeV$^2$,   
and in the second row the corresponding cut is $5$~GeV$^2$.   
For comparison, in the same kinematic region the inclusive diffractive    
cross section (for three flavors) is estimated to be $\sim 2.5$ nb,  
i.e. for the lower $k_t$ cut the combined $q\bar{q}$ and $q\bar{q}g$ jet rate  
amounts to approximately $8\%$ of the diffractive cross section. A look  
at the $k_t$ spectra indicates that the $q\bar{q}g$ jet rate   
strongly increases if we lower the cutoffs (entering a region where our  
formula requires a modification of the gluon structure function): this  
indicates that the largest part of the $q\bar{q}g$ cross section might come  
from an intermediate region where, in particular, the gluon $k_t$ is larger   
than the soft Pomeron scale but lower than the cutoff values used in our   
analysis. 
\par 
As in the case of diffractive production of two jets we expect the   
$q\bar{q}g$ jet cross section to rise in $1/x_{\fP}$; for the special case   
$\kf_2^2 \ll \kf_1^2$ such a behavior follows immediately from   
eq.~(4.5):   
$\sigma \sim \left[x_{\fP} g(x_{\fP},\mu^2)\right]^2$.  
\begin{figure}[htb]   
\begin{center}   
\epsfig{figure=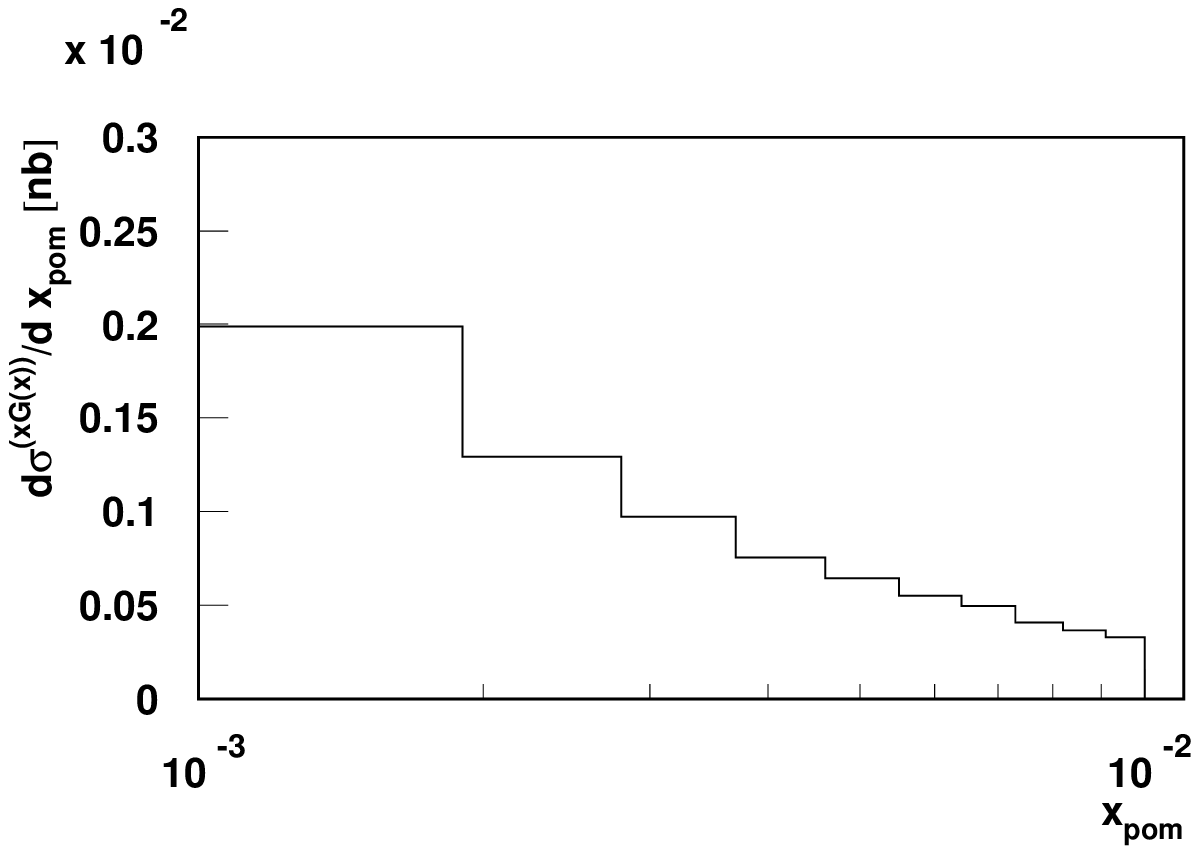}   
\end{center}   
\captive{The cross section as a function of $\xpom$ at fixed $\beta=0.1$.  
The cross section is integrated over $t$.   
\label{xpom}}   
\end{figure}   
In Fig.~\ref{xpom} we show the cross section as a function of $\xpom$ at fixed   
$\beta$. In the region $\beta=0.1$ (i.e. large diffractive mass) we expect    
our formula to be applicable. One can see the steep rise of the cross    
section at small $\xpom$, a   
similar behavior as observed already in the dijet case. This rise    
is in qualitative agreement with the   
approximation eq.~(4.5).  
\begin{figure}[htb]   
\begin{center}   
\epsfig{figure=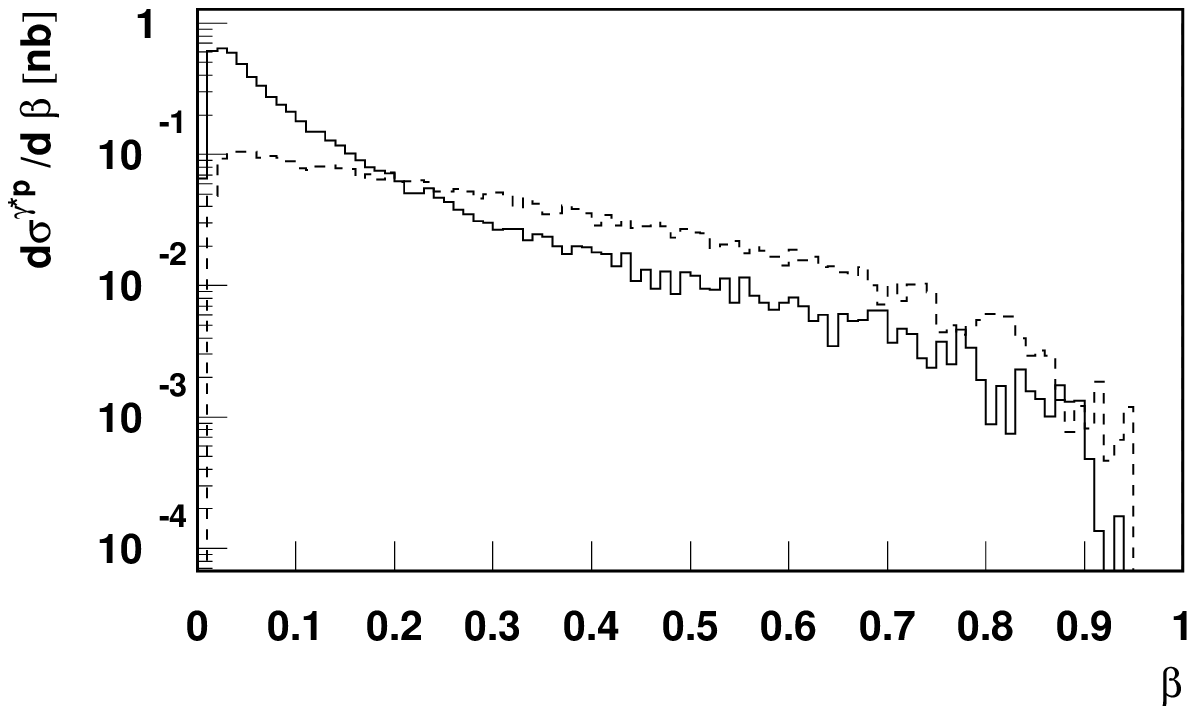}   
\end{center}   
\captive{Diffractive jet cross sections $d\sigma/d\beta$ at $t=0$  
 for $0.01 < y < 0.9$,   
$Q^2 > 5$, and $x_{\fP} < 0.01$. All $k_T^2$'s are restricted to be larger    
than $2$ GeV$^2$. The solid lines shows the $q\bar{q}g$   
final state, the dotted one the final state $q\bar{q}$.   
\label{beta}}   
\end{figure}   
As we have indicated in the introduction, we expect the $q\bar{q}g$ diffractive   
final state to dominate in the region of large diffractive masses (small   
$\beta$). In Fig.~\ref{beta} we show, for different $\beta$ bins, the    
integrated cross sections of the two diffractive processes $q\bar{q}$    
and $q\bar{q}g$ production. One sees   
that the $q\bar{q}g$ becomes dominant approximately for $\beta<0.2$.  
It is interesting to note that this $\beta$-value is consistent with the   
fit to the diffractive cross section in~\cite{BEKW}: this fit also suggests   
that the $q\bar{q}g$ contribution begins to dominate for $\beta \approx  
0.2$. In~\cite{BR}, the solution with a large $\gamma$-value is closer to  
our estimate than the small-$\gamma$ solution. 
\par
\begin{figure}[htb]   
\begin{center}   
\epsfig{figure=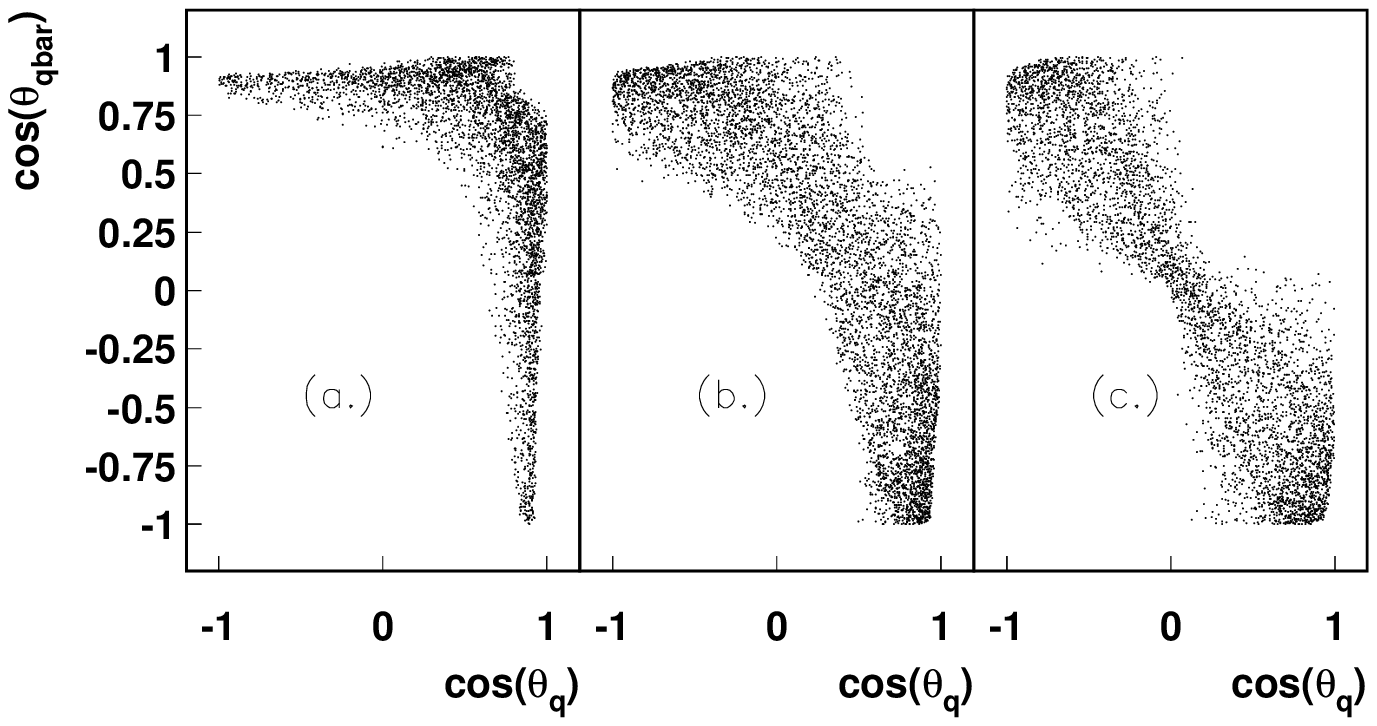}   
\end{center}   
\captive{The $\theta$ angle of the quark  versus the   
$\theta$ angle of the antiquark. ($a.$) for   
$-1 < \cos(\theta_{gluon})< -0.8$,    
($b.$) for $-0.6 < \cos(\theta_{gluon})< -0.4$, ($c.$) for    
$-0.2 < \cos (\theta_{gluon})< 0.0$,   
 at $\beta=0.1$ and $Q^2=50$ GeV$^2$ ($M^2_X=45$ GeV$^2$).   
\label{corr_theta}}   
\end{figure}

\begin{figure}[htb]   
\begin{center}   
\epsfig{figure=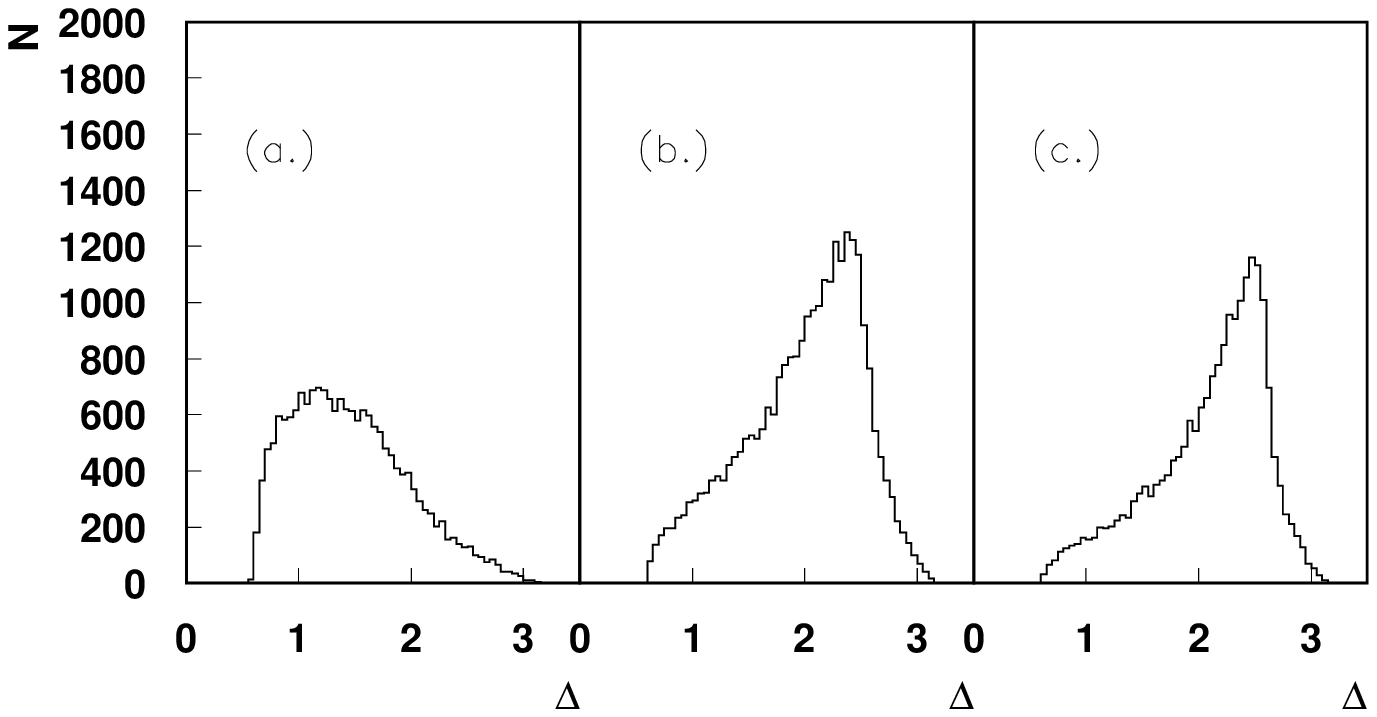,}   
\end{center}   
\captive{The opening angle $\Delta$ between the quarks.   
($a.$) for $-1 < \cos(\theta_{gluon})< -0.8$,    
($b.$) for $-0.6 < \cos(\theta_{gluon})< -0.4$, ($c.$) for    
$-0.2 < \cos (\theta_{gluon})< 0.0$,   
 at $\beta=0.1$ and $Q^2=50$ GeV$^2$ ($M^2_X=45$ GeV$^2$).   
\label{opening_angle}}   
\end{figure}

 
Next we study in more detail the specific properties of the $q\bar{q}g$    
system, in particular the angular distribution of the three jets.  
We begin with a scatter plot of the angles of the quarks, keeping the   
angle of the gluon fixed. Quarks angles are defined as angles between   
quark and photon directions. In Fig.~\ref{corr_theta} we show the results for   
three different regions of the angle $\theta_2$ of the gluon. According to   
eq.~(2.22), this angle is related to the transverse momentum $\kf_2$ of the   
gluon. At small gluon angle (direction of the gluon close to the   
proton (Pomeron) direction, Fig.~\ref{corr_theta}$a.$), in most of the events    
both quark and antiquark are moving not far from the photon direction    
(i.e. both their $\cos\theta$'s are close to 1).   
However, there are also events in which   
one of the quarks stays close to the photon direction, whereas the other one   
sticks out at a larger angle. At large angle of the gluon   
(gluon direction orthogonal to the photon-Pomeron beam axis,    
Fig.~\ref{corr_theta}$c.$), on the other hand, the quarks tend to move in    
opposite directions,   
one in the photon direction and the other in the proton direction.   
Correspondingly, at small gluon angle the opening angle $\Delta$ between the    
quarks prefers to be around $\pi/3$ (Fig.~\ref{opening_angle}$a.$), whereas for    
large gluon angle (Fig.~\ref{opening_angle}$c.$)     
$\Delta$ peaks at a value slightly above $2\pi/3$.  
$ $  
In all these discussions it should be made    
clear, that case ($c.$) lies at the edge of the region of validity of our    
approximation. A reliable calculation of this kinematic region requires   
a complete NLO calculation of diffractive $q\bar{q}$ and $q\bar{q}g$    
production, in particular virtual corrections to $q\bar{q}$ production and   
real gluon emission from quark lines. 
\par  
From these considerations we arrive at the following picture:   
\begin{itemize}   
\item At small transverse momenta of the gluon, we have mainly two-jet    
configurations where the quarks travel in approximately the same direction    
with an opening around $\pi/3$, carrying similar longitudinal momenta. But   
the two quarks can also appear in a more asymmetric configuration, where    
one of the quarks has a much larger longitudinal momentum compared to the    
other quark, and moves in approximately the direction opposite to the gluon.   
The quark with the smaller longitudinal momentum then sticks out from the   
$z$-axis.    
\item   
In the other extreme, where the gluon is emitted perpendicular to the $z$-axis,   
we observe a $'Mercedes-star"$ like configuration, in which both quarks carry    
longitudinal momenta of similar value, but in opposite direction,    
and the cosine of the opening angle lies between two and three.   
\item   
At medium transverse momentum $\kf_2$ of the gluon, all configurations    
discussed above, are possible.   
\end{itemize}   
  
\section{Conclusions}     
In this paper we have presented an analysis of hard diffractive $q\bar{q}g$    
jets   
with large transverse momenta. Our study is motivated by the observation   
that in DIS diffractive dissociation, the Pomeron intercept lies above the   
value seen in purely hadronic interactions: this suggests that   
the diffractive final state contains a rather large `hard` component.   
A natural candidate for hard final states are jets. The investigation    
of diffractive $q\bar{q}g$ production presents a generalization   
of diffractive dijet production~\cite{BLW}. Our calculation is restricted   
to the low-$\beta$ region (large diffractive masses), and we have been   
working in the leading-log $1/\beta$ approximation. At the same time, the   
transverse momenta of the outgoing partons are not restricted by a strong   
ordering requirement. In the low-$\beta$ region the emitted gluon carries   
a small momentum fraction of the incoming photon, i.e. in rapidity it is   
closer to the proton than the two quarks (Fig.~\ref{notation}).\\ \\   
The analytic expression for the cross section formula   
in impact parameter space is remarkably simple.   
It illustrates that the concept of a photon wave function holds even   
beyond the approximation, where the transverse momentum of the emitted gluon    
is much softer than the quarks. For special kinematic limits we have obtained   
even simpler expression for the cross section formulae.\\ \\   
In a first exploratory numerical study we have calculated jet cross sections.   
Depending upon the lower limit of the transverse momenta, the combined    
$q\bar{q}$ and $q\bar{q}g$ jet cross section can be as large as $8\%$ of the   
inclusive diffractive cross section.   
We have also looked into the spatial distribution of the three-parton final    
state. If (in the $\gamma^* - \PO$ CM-system) the gluon jet is close to the   
Pomeron direction, the quark - antiquark pair mainly forms a single (although    
somewhat broader) jet opposite to the gluon jet. In the extreme case where   
the emitted gluon is perpendicular to the photon - Pomeron axis, a clean    
three-jet configuration emerges. However, in this region the cross section will   
be small. For a diffractive jet analysis, it is important to keep in mind that   
quite a large fraction of the $q\bar{q}g$ final states will appear as a   
two-jet configuration, and it will be difficult to separate them from   
the $q\bar{q}$ dijet final states. \\ \\  
Experimentally the analysis of two or three jet events is not easy.  
A more promising way may be a thrust or $E_T$ analysis.  
To this end, it would be very desirable to have a complete NLO calculation   
of diffractive final states with two or three jets.   
\section*{Acknowledgement}
Part of this work was done while one of us (J.B.) was visiting the 
Fermilab Theory Division. He gratefully acknowledges their hospitality.

\end{document}